\documentclass[aps,pra,twocolumn,showpacs]{revtex4}

\usepackage{graphicx}
\usepackage{color}
\usepackage{bm}
\usepackage{amsmath, amsthm, amssymb,mathrsfs}
\usepackage{csquotes}
\usepackage[colorlinks=true, citecolor=blue,allcolors=blue]{hyperref}

\begin{document}

\title{Photo\-electron momentum distributions in the strong-field ionization of 
atomic hydrogen by few-cycle elliptically polarized optical pulses}

\author{Nicolas Douguet$^1$ and Klaus Bartschat$^2$}

\affiliation{$^1$Department of Physics, Kennesaw State University, Marietta, 30060, USA\\
$^2$Department of Physics and Astronomy, Drake University, Des Moines, Iowa 50311, USA}

\date{\today}

\pacs{32.80.Rm, 32.80.Fb, 32.80.Qk, 32.90.+a}

\begin{abstract}
We investigate the strong-field ionization of atomic hydrogen in a few-cycle elliptically polarized infrared pulse by solving 
the time-dependent Schr\"{o}dinger equation. The dependence of the photo\-electron momentum distribution
on the pulse intensity, ellipticity, length, envelope, and carrier envelope phase is analyzed. In particular, we explain the variation 
of the electron offset angle with asymptotic electron energy through the combined action of the field and the Coulomb potential, and 
demonstrate that low-ellipticity pulses make it possible to access the electron release time.
\end{abstract}

\maketitle

\section{Introduction}
\label{sec:1}
The strong-field ionization (SFI) of atoms and molecules by intense femtosecond elliptically-polarized infrared (IR) pulses has
attracted considerable interest in recent years. Besides leading to the generation of circularly polarized bi-harmonics   
\cite{Ferre14,photonics4020031,Reich16,PhysRevLett.117.133201,PhysRevA.52.2262,PhysRevA.62.011403},
elliptical pulses have improved our understanding of fundamental strong-field processes, e.g.,
frustrated tunneling ionization~\cite{PhysRevA.102.013116} and multi\-photon circular 
dichroism~\cite{PhysRevLett.126.023201,PhysRevA.103.053125}. In particular, 
the attoclock set\-ups \cite{Torlina15,Landsman14,Zimmermann16,PhysRevLett.119.023201,Satya18,Peipei18,Peipei19,PhysRevLett.127.273201}, 
using either one-color or two-color elliptical pulses, have allowed us 
to access the phase \cite{Peipei18}, tunneling time \cite{Satya18}, Wigner time delay \cite{Trabert21}, and 
the overall temporal evolution of an emitted electron wavepacket~\cite{Peipei18,Wang14} in an intense IR field.

Attoclock set\-ups, which have been used on several atoms, such as hydrogen~\cite{Satya18,PhysRevLett.127.273201}, 
helium~\cite{Eckle1525,Pfeiffer12}, 
argon~\cite{Pfeiffer12,PhysRevLett.119.023201}, and krypton~\cite{PhysRevLett.119.023201}, and recently on the 
hydrogen molecule \cite{Hanus18,Serov19},
propose to correlate measurable observables in the photo\-electron momentum distribution (PMD) to the 
ultrafast dynamics of the electron wavepacket in an intense laser field. Of critical importance is the 
so-called {\it offset angle} \cite{Landsman14,Satya18,PhysRevLett.127.273201} of the asymptotic photo\-electron momentum
with respect to the asymptotic momentum of a classically free electron released with zero velocity at the moment of maximum field strength.

The relation between the offset angle and the tunneling time of the electron under a potential barrier 
remains a central subject that has triggered vigorous debates and 
controversies~\cite{Torlina15,Satya18,Satya2020,Kheifets2020,PhysRevLett.119.023201},
while a complete understanding of its dependence with the pulse characteristics is 
still the focus of current efforts~\cite{PhysRevLett.127.273201}. 
In neutral targets, a complication arises from the effect of the Coulomb field, which ultimately induces an 
additional deflection of the electron momentum. 
The effect would, however, disappear in the strong-field ionization of a negative ion such as F$^-$, 
as we originally proposed in \cite{Douguet19}.  This scenario was recently
revisited in a fully correlated approach by Armstrong {\it et al.}~\cite{Armstrong20} 
using the $R$-matrix with time-dependence (RMT) method.
As a final complication, the offset angle in a Coulomb field actually depends on the definition of this angle itself.

Experimentally, the PMD has been measured in atomic hydrogen~\cite{Satya18} for a nearly circular (ellipticity $\epsilon=0.87$) 
800-nm pulse with a FWHM of $\approx 6\,$fs
for peak intensities varying from $1.65$ to $3.9\times10^{14}\,$W/cm$^2$. The role of the elliptical field 
is to define a reference axis on which the $E$-field takes its 
maximum value.  This reference was necessary since
the carrier-envelope phase (CEP) was not stabilized in the latter experiment. 

In the near future, it is reasonable to expect that even shorter and better-defined pulses will be generated, which
will enable the exploration of the electron dynamics in more detail. As one is able to tailor 
few-cycle pulses, our understanding of the electron dynamics under different pulse conditions becomes essential.
Even for a system as simple as atomic hydrogen, the interpretation of the PMD for different 
pulse characteristics is far from being straightforward. 
The shape of the pulse, 
characterized by the temporal positions of its maxima, as well as the laser intensity, 
can result in drastic changes in the electron dynamics, which are directly encoded in the PMD. 

As numerous theoretical and experimental studies have been reported under various conditions, it is desirable 
to systematically analyze the principal features of the PMD in atomic hydrogen for different pulse parameters 
(envelope shape, CEP, ellipticity, peak intensity, and pulse length)
and to elucidate the origin of these features when possible. Below, we also emphasize
particular choices of pulses that lead to the most salient physical effects with the aim of motivating future
experimental and theoretical studies. In particular, we explain the drift of the offset angle with electron energy \cite{Trabert21} 
and reveal its connection with the electron release time.

Although experiments on 
atomic hydrogen are challenging (but certainly possible~\cite{Satya18,Trabert21}), we use this target, since
this is the problem that can be solved numerically with high accuracy and without multi\-electron effects. The latter generally
need to be approximated in some way, thereby potentially affecting the conclusions. 

This manuscript is organized as follows. In Sec.~\ref{sec:2}, we describe the theoretical approach and the framework of the 
performed simulations. We then present and discuss the results for various pulse parameters in Sec.~\ref{sec:3}.
We set the stage by describing general features of the PMD before presenting our main result in 
Sec.~\ref{sec:ellipticity_dependence}, namely the possibility to effectively select the release time of the   
electron by employing suitable short pulses with low ellipticity.
Section~\ref{sec:4} is devoted to our conclusions.

Unless indicated otherwise, atomic units are used throughout the manuscript. 

\section{Theoretical approach }
\label{sec:2}
\subsection{Numerical simulations}
The time-dependent Schr\"{o}dinger equation (TDSE) for the wavefunction $\Psi(\bm r,t)$, 
where $\bm r$ is the vector position of the electron, takes the form
\begin{equation}
i\frac{\partial \Psi(\bm r,t)}{\partial t}=\left[H_0+H_{F}(t)\right]\Psi(\bm r,t).
\end{equation}
Here $H_0$ is the field-free hamiltonian of atomic hydrogen. The electron-field 
inter\-action $H_F(t)$ in the dipole approximation is expressed, either in the
length gauge (LG) or in the velocity gauge (VG), as
\begin{equation}
H_{F}(t)=\left\{\begin{array}{l}
{\bm E}(t)\cdot \bm r~~{\rm (LG)};\\
{\bm A}(t)\cdot \bm p~~{\rm (VG)}.
\label{eq:gauge}
\end{array}
\right.
\end{equation}
The TDSE is solved by expanding the wavefunction in partial waves as
\begin{equation}
\Psi(\bm r,t)=\sum_{\ell=0}^{\ell_{max}}\sum_{m=-\ell}^{m=\ell}\xi_{\ell m}(r,t)Y_{\ell m}(\theta,\phi),
\end{equation}
where $(r,\theta,\phi)$ are the electron spherical coordinates, $\ell$ and $m$ are the electron angular momentum 
and its projection on the $z$-axis, respectively,
while $Y_{\ell m}$ are spherical harmonics. 

We then propagate the initial state by splitting the time-evolution operator as
\begin{equation}
e^{-iH(t+\frac{\Delta t}{2})\Delta t}= e^{-iH_0\frac{\Delta t}{2}}e^{-iH_F(t+\frac{\Delta t}{2})dt}e^{-iH_0\frac{\Delta t}{2}},
 \end{equation}
where $H(t)=H_0+H_F(t)$ is the total hamiltonian of the system.
The Crank-Nicolson propagation scheme is used to expand the time-evolution operator. For the field-electron 
inter\-action, for instance, we obtain
\begin{equation}
\label{eq:interaction_ham}
e^{-iH_F(t+\frac{\Delta t}{2})\Delta t}\approx\frac{{\bf I}-i\frac{\Delta t}{2}H_F(t+\frac{\Delta t}{2})}{{\bf I}+i\frac{\Delta t}{2}H_F(t+\frac{\Delta t}{2})},
 \end{equation}
where ${\bf I}$ is the identity operator. We employ a radial grid and a three-point formula to compute the electron kinetic energy. 
The time-evolution due to the field-free hamiltonian, $\exp[-iH_0\Delta t/2]$,
is computed by solving a tridiagonal system of linear equations on the radial grid. We use a grid with radial step $0.2$ 
and a box of $2,000$.  
The time step is varied depending on the intensity considered and we always assess
that the final norm of the wave function is unity to a precision of a least $10^{-8}$.

The time evolution of the field-electron inter\-action
is obtained by expanding the denominator in Eq.~(\ref{eq:interaction_ham}) 
in a Taylor series. We obtained good efficiency using a fourth-order expansion
and taking advantage of the sparsity of $H_F$ in both gauges. At the end of the pulse, we project the wavefunction onto 
momentum-normalized Coulomb functions $\phi_{\bm p}(\bm r)$,
with asymptotic momentum $\bm p$, to obtain the asymptotic PMD as $\mathcal P(\bm p)=|\langle \phi_{\bm p}(\bm r)|\Psi(\bm r,t)\rangle|^2$.
In the following discussion of elliptically polarized light, we restrict our study to the PMD in the polarization plane, i.e., we fix $p_z=0$. All PMDs presented in this study are expressed in a.u.$^{-3}$.

\subsection{Pulse characteristics}

We consider $N$-cycle plane-polarized infrared pulses of period $T$ defined in a time interval $t_i\le t\le t_f$, with $t_i=0$ and $t_f=NT$, 
by a vector potential propagating along the $z$~axis and written as
\begin{equation}
\label{eq:vector_potential}
{\bm A}(t)=-\frac{E_0}{\omega}\frac{f(t)}{\sqrt{1+\epsilon^2}}\left[\epsilon\sin(\omega t +\phi) 
{\hat{\bm e}_x} -\cos(\omega t +\phi){\hat{\bm e}_y}\right].
\end{equation}
In the above expression, ${\hat{{\bm e}}_x}$ and ${\hat{{\bm e}}_y}$ are unit vectors along the $x$ 
and $y$ axes, respectively, $\omega$ is 
the central laser frequency, $\epsilon$ is the field ellipticity,
and~$\phi$ is the CEP. The pulse envelope $f(t)$, which has a maximum value of unity, is chosen to be 
either $f(t)=\sin^2(\omega t/2N)$ ($\sin^2$ envelope), 
$f(t)=\sin^4(\omega t/2N)$ ($\sin^4$  envelope), or a gaussian envelope $f(t)=\exp[-4\ln2(t-t_m)^2/t_m^2]$, where $t_m=NT/2$ is the time 
of maximum value of the envelope, $f(t_m)=1$. 
For the latter, a small correction at the beginning and the end of the pulse is applied to ensure a zero field at the same
times as for the $\sin^{2p}$ envelopes.

\begin{figure}[thbp]
$~~$\includegraphics[width=6.5cm]{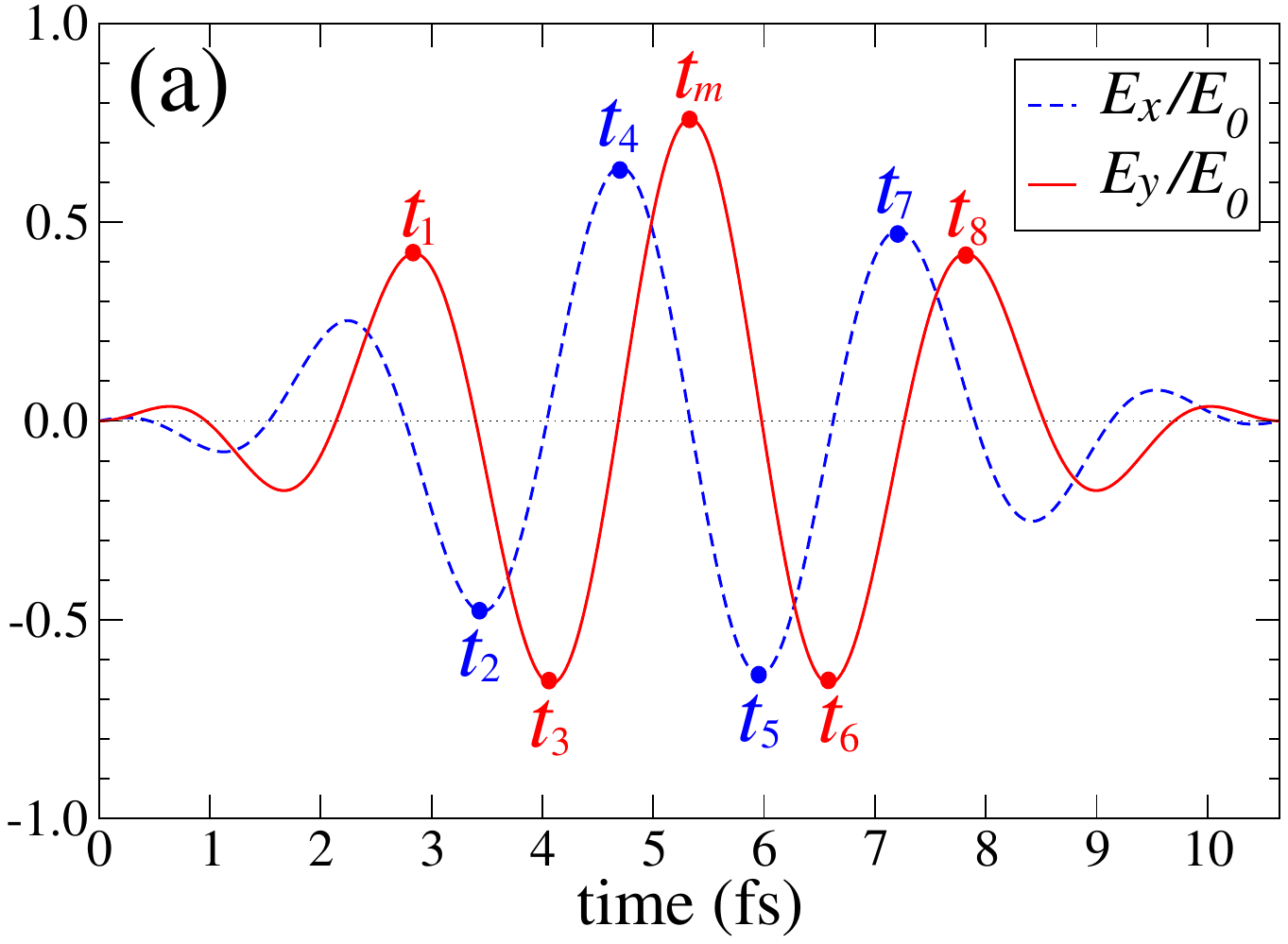}\\
\includegraphics[width=6.9cm]{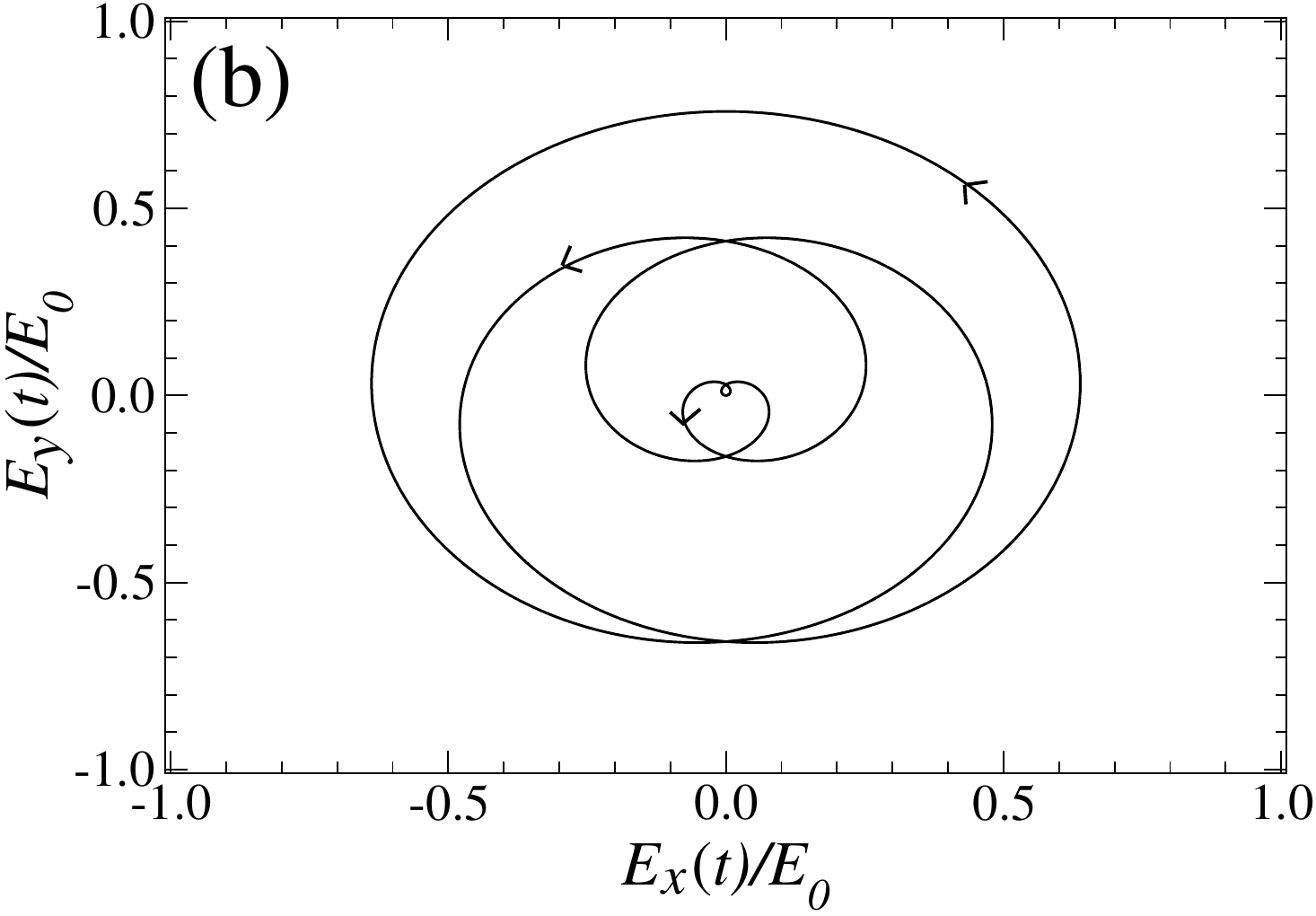}
 \caption{Example of an elliptical pulse employed in this study, with $\epsilon=0.87$, $N=4$, sin$^2$ envelope, and $\phi=90^\circ$. 
The field components are drawn (a) as a function of time, and (b) as a trajectory in the polarization plane. }  
  \label{fig:1}
\end{figure}

The resulting electric field ${\bm E}(t)=-d{\bm A}(t)/dt$ has positive helicity with a typical form 
shown in Fig.~\ref{fig:1}. For all numerical simulations
presented in this study, we fix the laser frequency at $\omega=0.057\,$a.u., corresponding to a central wavelength of $800\,$nm. 
We define the cycle-averaged intensity~$I$ 
around the peak of the envelope, such that $I=3.51\,E_0^2\times 10^{16}\,$W/cm$^2$, with $E_0$ expressed in atomic units.  
With this definition, 
the intensity becomes independent on the ellipticity~$\epsilon$.

Figure~\ref{fig:1} depicts a pulse taking its maximum value along the vertical axis ($\phi=90^\circ$) 
at the peak of the pulse envelope, 
where one expects the tunneling rate to be the largest.
Already for a 4-cycle pulse, several extrema along each axis contribute to the final electron signal, as they correspond to instants 
at which the electron can be released with a significant ionization rate, thereby leading to a complex ionization mechanism reflected in the PMD.

The pulse shown in Fig.~\ref{fig:1} fulfills two important criteria~\cite{PhysRevA.91.053404}.
First, by defining the vector potential ${\bm A}(t)$ instead of the electric field ${\bm E}(t)$,
we ensure that ${\bm A}(t_i)=0$ and ${\bm A}(t_f)=0$: a {\it sine qua non} to produce a realistic pulse. 
Starting directly from the definition of the electric field 
could lead to erroneous results in the velocity gauge for sufficiently short pulses when, 
as is generally the case except in very special choices, $\int_{t_i}^{t_f} {\bm E}(t)\ne 0$ and thus ${\bm A}(t_f)\ne 0$. 

In addition, the pulse depicted in Fig.~\ref{fig:1}
induces a vanishing net displacement of the free electron in the field, i.e., $\Delta \bm r=\int_{t_i}^{t_f} {\bm A}(t)=0$. 
Although a zero electron net displacement 
is not a strict requirement of Maxwell's equations (e.g., single-cycle THz pulses~\cite{PhysRevLett.112.143006} have finite net displacement),
it can be shown \cite{Joachain2011} that the field created by any laser cavity, i.e., such that $A(\omega)=0$ for $\omega\le\omega_{\rm min}$,
with $\omega_{\rm min}$ denoting the minimum cut-off angular frequency of a laser oscillator, leads to $\Delta \bm r=0$. 
In addition, fields with $\Delta \bm r\ne0$ can lead to
theoretical and experimental difficulties, e.g., a strong asymmetry in the Kramers-Hennerberger gauge or the fact that 
electrons do not remain in the laser focus.
A discussion of some of the issues can be found, for example, in~\cite{PhysRevA.90.043401}.
Nevertheless, pulses with non\-zero displacement might have interesting properties to control 
the spatial motion of the electronic density in large biomolecules, and experimental protocols to measure electron 
displacements have recently been designed \cite{Xiao19}.

It is somewhat surprising that the net displacement of the short elliptical pulses employed in many theoretical 
studies has hardly been discussed. 
Hence we pause to briefly comment on this specific point. 
First, a nonzero net displacement is due to a combination of the field envelope 
and the CEP. For light that is linearly polarized along the $z$~axis, choosing a CEP such that the pulse is 
antisymmetric (odd) with respect to $t_m$ will ultimately 
result in $\Delta z=0$. For elliptical light, however, the pulse components of the form (\ref{eq:vector_potential}) 
cannot be simultaneously 
anti\-symmetric along both the $x$~and $y$~components. Consequently, the net displacement cannot vanish by symmetry. 
In the case of a gaussian pulse 
with fixed peak intensity, the displacement decreases quickly as a function of the FWHM of the pulse. For instance, at an intensity 
of $2\times 10^{14}\,$W/cm$^2$ and $\phi=90^\circ$,
the displacement on the $x$ axis for 800-nm light is $\Delta x=4.7\,$a.u.\ for FWHM~$=2\,$fs, but only $\Delta x=0.04\,$a.u. for FWHM~$=3\,$fs. 
Note that at long wavelengths, such as $\lambda=2000$~nm,
the displacement for such short gaussian pulses can become $\Delta x>100$~a.u.
On the other hand, considering $f(t)=\sin^{2p}(\omega t/2N)$ as the envelope, it can readily be shown 
that $\Delta \bm r=0$ for pulses with $N\ge p$ cycles.
For $N<p$, however, the net displacement $\Delta \bm r$ can take anomalously large values. 
We note that the $\sin^4$ envelope with $N=2$ 
(often characterized as a nearly single-cycle pulse in the literature), which has been widely used 
in recent theoretical works such as~\cite{Torlina15}, 
leads to a displacement as large as $\Delta x=12.5\,$a.u. 
at an intensity of $2\times 10^{14}\,$W/cm$^2$ for $\phi=90^\circ$.

\section{Results}
\label{sec:3}
We start this section by giving a general description of the PMD, and we also assess the accuracy 
of our calculations by analyzing the gauge invariance.
We then analyze and interpret  the PMD for different pulse characteristics, varying one parameter at a time.  
Specifically, we consider variations of the envelope shape, CEP, pulse intensity, ellipticity, 
and pulse length. 

While we restate at times some familiar considerations for completeness of the discussion, we draw special attention to
Secs.~\ref{sec:intensity dependence} and ~\ref{sec:ellipticity_dependence}, where we analyze in detail the intensity and ellipticity dependence of the PMD.
As a key result, we demonstrate how
short and intense optical pulses with relatively low ellipticity ($0.4\le \epsilon\le0.6$) provide important information on various aspects of the electron 
tunneling dynamics in strong fields, and can actually serve as selectors of the electron release time.

\subsection{General description of the PMD}
\label{sec:3:a}

We first consider our benchmark pulse in Fig.~\ref{fig:1}, as it is close to a pulse that ({\it i})~can be generated 
experimentally, and ({\it ii})~enables the extraction of a number of interesting physical effects.
This pulse has a $\sin^2$ envelope, an ellipticity $\epsilon=0.87$, and a CEP $\phi=90^\circ$.  It results in 
the maximum ionization probability along the $y$~axis, 
similarly to what has been considered in recent experimental and theoretical works~\cite{Satya18,PhysRevA.93.023425}. 

\begin{figure}[tbp]
\includegraphics[width=8.cm]{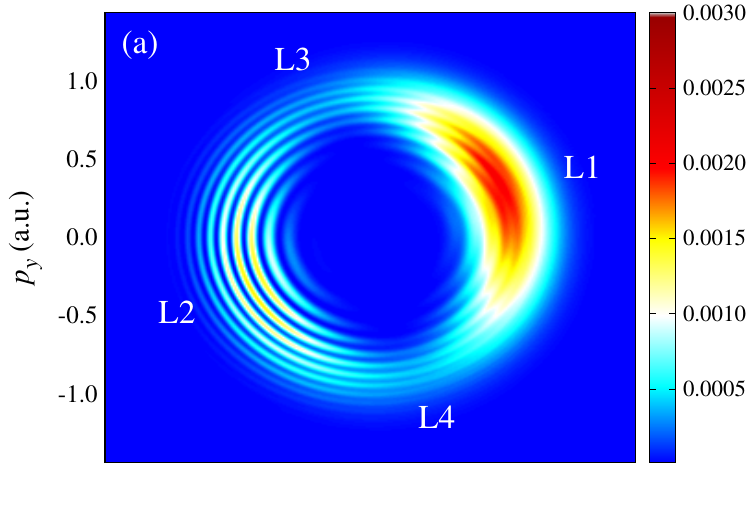}\\
\vspace{-0.5cm}
\includegraphics[width=8.cm]{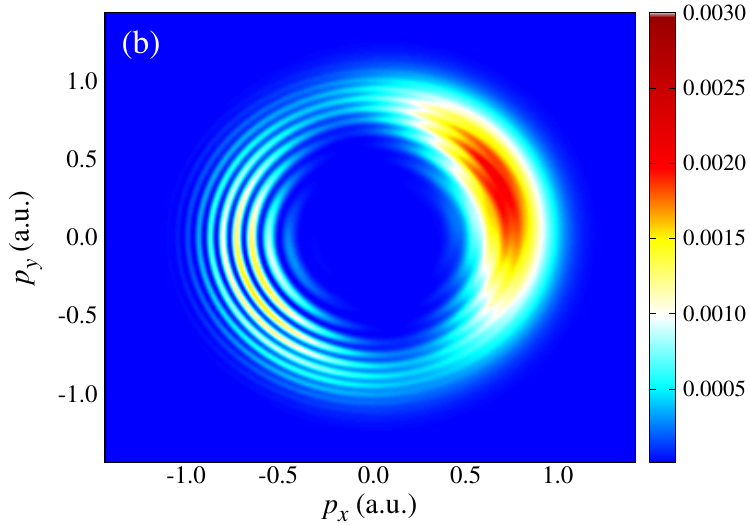}
\caption{PMD for pulse parameters $\epsilon=0.87$, $N=4$, $I=10^{14}\,$W/cm$^{2}$, sin$^2$ envelope, and $\phi=90^\circ$ calculated
using (a)~the length gauge and (b)~the velocity gauge. The features L1$-$L4 are explained in the text. \label{fig:2}}  
\end{figure}

Using the pulse described above, we first choose a peak intensity of $10^{14}\,$W/cm$^2$ and compare 
the PMDs obtained in the length and velocity gauges 
to assess the accuracy of our calculations. The results are presented in Fig.~\ref{fig:2}, where  
we use $\ell_{max}=100$ in the length gauge and $\ell_{max}=60$ in the velocity gauge.
Excellent agreement is observed between the results from the two calculations.
The velocity gauge in this intensity regime is known to be drastically more efficient than the length gauge~\cite{CorLam1996,PhysRevA.81.043408}.  
Therefore, we employed this gauge in all the remaining calculations reported in this paper.

\begin{figure*}[htbp]
 \begin{center}
 $
  \begin{array}{ccc}
  \vspace{-0.47cm}
  \includegraphics[width=5.8cm]{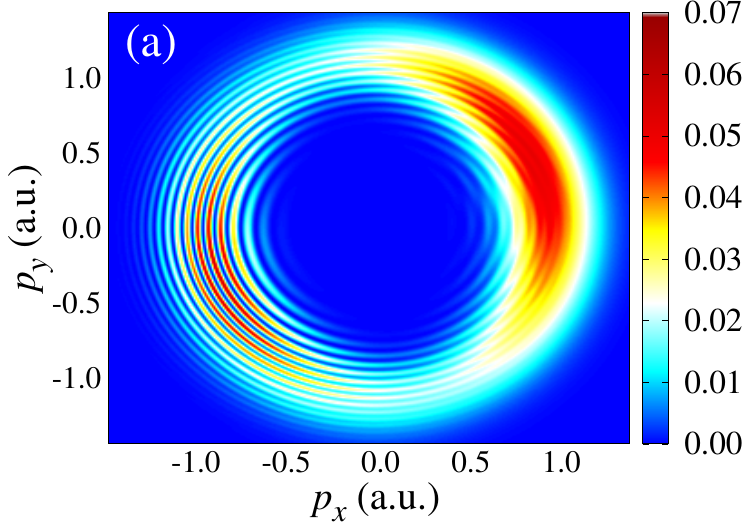} 
  &\includegraphics[width=5.8cm]{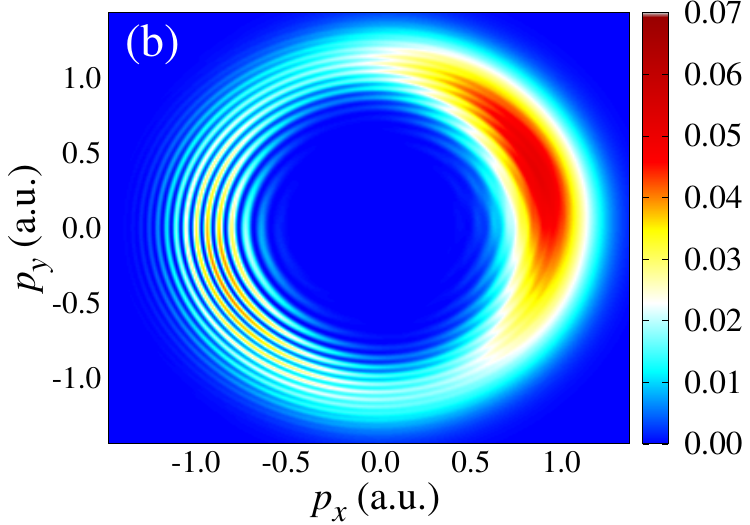} &\includegraphics[width=5.8cm]{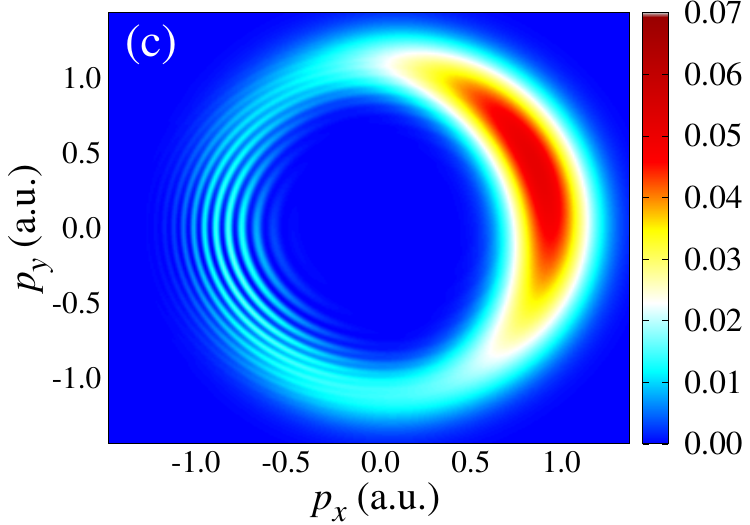}
 \end{array}
$
\end{center}
\caption{PMDs for pulse with $\epsilon=0.87$, $N=4$, $I=2\times 10^{14}\,$W/cm$^{2}$, 
$\phi=90^\circ$, with \hbox{(a) sin$^2$}, \hbox{(b) gaussian }, and \hbox{(c) sin$^4$} envelopes.}  
\label{fig:3}
\end{figure*}

The PMD in Fig.~\ref{fig:2} exhibits characteristic features that deserve some general comments. First, the PMD has 
two principal lobes, L1 and L2 (see Fig.~\ref{fig:2}a), which share 
an approximate symmetry axis shifted with respect to the horizontal axis. If the electron is released along the $y$~axis 
at the moment $t_m$, when the field assumes its largest value, we know from Eq.~(\ref{eq:vector_potential}) 
that $\bm A(t_m)=-E_0\epsilon/\omega\sqrt{1+\epsilon^2}\hat{e}_x$.
As a result, if the electron is released with zero initial speed and subsequently inter\-acts with the field only,
its asymptotic momentum ${\bm p}=-\bm A(t_m)$ should be along the positive $x$~axis (since $\bm A(t_m)\cdot\hat{e}_x<0$). 
However, the electron is also strongly deflected by the Coulomb 
force, which results in an offset of the asymptotic momentum. Below, we do not restrict our discussion to the 
variation of the offset angle with intensity \cite{Torlina15,Landsman14,Zimmermann16,PhysRevLett.119.023201,Satya18,Bray18},
but provide a more general theoretical description of the main features of the PMD. The dependence 
of the offset angle on the laser intensity, laser ellipticity, and the
photoelectron energy, will be discussed in more detail in Secs. \ref{sec:intensity dependence} and \ref{sec:ellipticity_dependence}.

For a very short pulse, such as $N=2$, we expect a single lobe 
in the PMD, because the electron is released almost exclusively when the field takes its maximum value at $t_m$. 
In the case of the $4$-cycle pulse of Fig.~\ref{fig:1}, however, there exist
two times, $t_3$ and $t_6$, located symmetrically with respect to $t_m$, which correspond to large negative values of the field 
component along the $y$~axis.  
The lobe L1 is thus primarily due to tunneling at $t_m$, with exit position $y<0$, 
whereas the lobe L2 is mostly due to tunneling at the two symmetric moments, $t_3$ and $t_6$, with exit positions $y>0$. 
However, given that the tunneling rate at $t_m$ is significantly larger than at $t_3$ and $t_6$ due to the exponential 
dependence of the tunneling rate on the field strength \cite{ADK86,Tong02}, 
the signal at L1 remains overall the largest.

Besides the signal strength, the two lobes also exhibit noticeable differences in shape. 
Because the lobe L2 results from two nearly identical electron bursts produced
at $t_3$ and $t_6$, separated by one IR period, it produces a high-contrast periodic series
of maxima and minima in momentum space at energies such that $E=n\hbar\omega-I_p-U_p$, 
where $n$ is the total number of absorbed photons, $I_p$ denotes the ionization potential, 
and $U_p=E_0^2/4\omega^2$ is the ponderomotive energy.
The series of peaks is convoluted by the energy uncertainty of the bursts and centered at their mean energy. On the other hand, 
the lobe L1 can be considered to be formed from three bursts, at $t_1$, $t_m$, and $t_8$. 
Because the electron burst produced at $t_m$ is significantly larger than those at 
$t_1$ and $t_8$, the PMD is dominated by a strong background with a superimposed series of 
maxima and minima with poor contrast. The repetition
of similar tunneling events toward a particular asymptotic direction, therefore, results in 
high-contrast fringes in the PMD. Furthermore, the energy distribution is different in each lobe, i.e.,
L2 has a stronger signal at low and high energies when compared to L1, because the electron detected 
at L2 is released either at an earlier ($t_3$) or a later time ($t_6$) than $t_m$,
thereby having, respectively, more or less probability to absorb photons. 

Using previous considerations, one can readily understand 
the features at L3 and L4, which are located at $90^\circ$ relative to the symmetry axis formed by L1 and L2 (see Fig.~\ref{fig:2}). 
This signal results from tunneling at $t_2$ and $t_5$ for asymptotic momentum $p_y>0$,
and at $t_4$ and $t_7$ for asymptotic momentum $p_y<0$. Since the tunneling events occur at equivalent field 
strengths along the $x$~axis, except for the time ordering, their momentum distributions
are similar, as seen in Fig.~\ref{fig:2}. Finally, because the electric field is similar at $t_2$ 
and $t_5$, as well as at $t_4$ and $t_7$, we see a relatively good contrast of the fringes at L3 and L4.  
As expected, however, it is not as high as for L2. 

Note that the deflection angles of the L1 and L2 lobes are nearly the same. This is consistent with the fact that the 
angle is mainly due to the long-range Coulomb inter\-action rather than the field-dependent tunneling process. 
The small difference between the angular offsets of L1 and L2 is due 
to the fact that ({\it i}) an electron released in the field at different locations and times inter\-acts for different time 
intervals with the combined Coulomb and electro\-magnetic fields, and ({\it ii})
that the vector potential at field extrema other than $t_m$ is not precisely aligned with the $x$ axis. 
The former effect explains
the differences in the asymptotic deflection angles at the various fringes, which was recently 
considered in~\cite{PhysRevLett.127.273201}.
For the large ellipticity considered here, the offset angle at the maximum signal of the individual fringes 
 is only weakly dependent on the fringe order. It will be shown, however, that this dependence increases 
 at smaller ellipticity (see Sec.~\ref{sec:ellipticity_dependence}).

In the following subsections, we consider the dependence of the PMD on the CEP, envelope shape, ellipticity, 
intensity, and pulse length. 
Varying these parameters independently, we keep the pulse presented in Fig.~\ref{fig:1} as our benchmark 
but increase its peak intensity to $I=2\times10^{14}\,$W/cm$^2$ to ensure that we are truly in the tunneling regime, i.e., 
with $\gamma\le1$, where $\gamma=\sqrt{I_p/2U_p}$ is the Keldysh parameter~\cite{Keldysh1965}. At $I=2\times10^{14}\,$W/cm$^2$, 
for instance, $\gamma \approx 0.75$ for atomic hydrogen at a wavelength of $\lambda = 800\,$nm. Although we do not 
include them in the figures below, we keep the labels of the lobes introduced in Fig.~\ref{fig:2}(a) to 
facilitate our discussion.

\begin{figure*}[htbp]
 \begin{center}
 $
  \begin{array}{c}
  \includegraphics[width=18.cm]{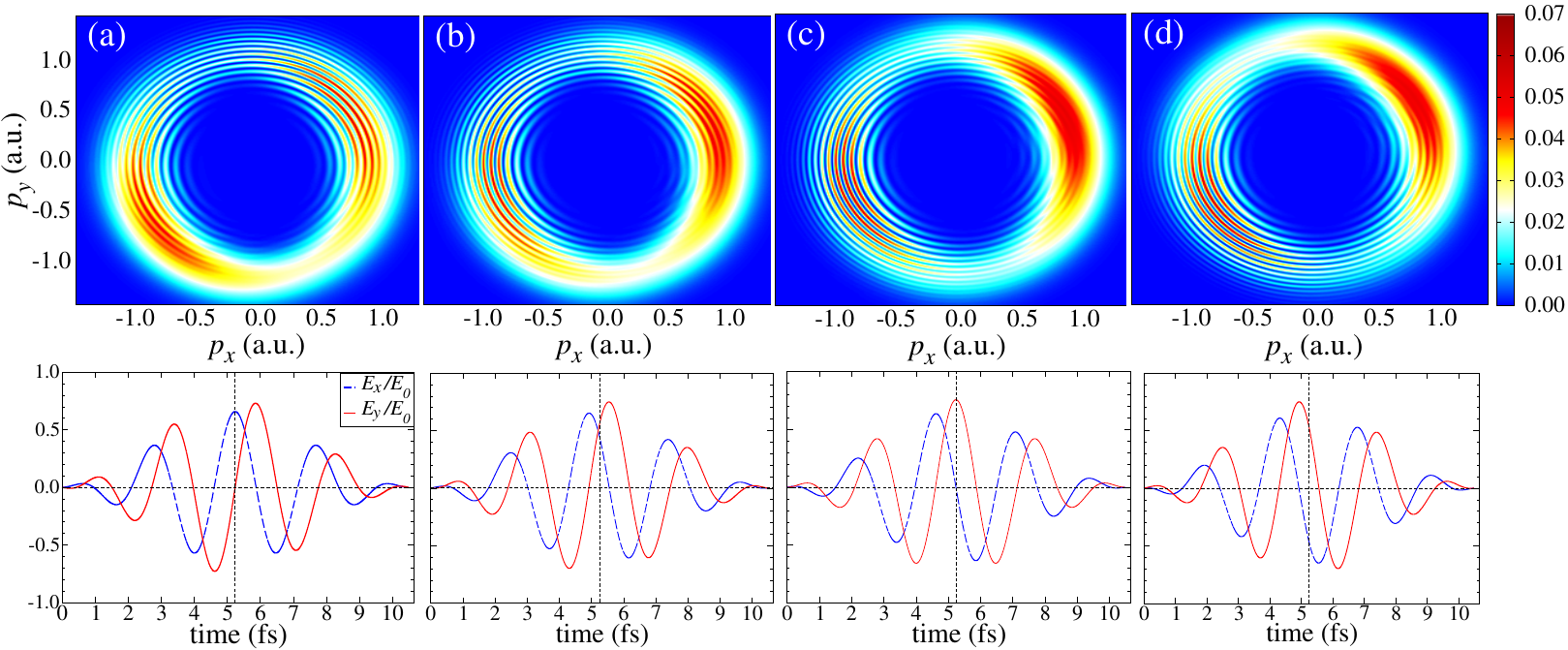} 
    \end{array}
$
\end{center}
\caption{PMDs (top) and associated pulses (bottom) for pulse parameters $\epsilon=0.87$, 
$N=4$, $I=2\times 10^{14}\,$W/cm$^{2}$, sin$^2$  envelope, and 
a CEP chosen as (a) $\phi=0^\circ$, (b) $\phi=45^\circ$, (c) $\phi=90^\circ$, and (d) $\phi=135^\circ$.
The horizontal and vertical axes in the bottom panels, respectively, indicate
zero electric field and the time $t_m$ of the maximum pulse envelope.}
\label{fig:4}
\end{figure*}

\subsection{Envelope dependence}
We first consider the dependence of the PMD on the choice of envelope function by employing our benchmark pulse, 
either with a gaussian, $\sin^2$, or $\sin^4$ envelope. 
Since the gaussian envelope has, in principle, an infinite number of cycles, we reduce it to $N=4$ cycles by adding
a small linear correction at the edge of the pulse to compare results at the same number $N$ of cycles. 
We also recall that the gaussian pulse does not guarantee a vanishing displacement. Nevertheless, we
use it for illustrative purposes, since for the parameters chosen here the displacement is sufficiently small
for meaningful conclusions to be drawn.

Our results, displayed in Fig.~\ref{fig:3}, reveal a weak dependence on the pulse envelope. The general trend can 
be explained by considering that the envelope width  
decreases gradually from $\sin^2$, to gaussian, to $\sin^4$ envelopes, thereby leading to a decreasing contrast of the 
fringes at L1 as the tunneling rates at 
$t_1$ and $t_8$ become much smaller than at $t_m$. Similarly, the signal strength at L2 decreases with pulse 
width due to diminishing tunneling rates at $t_3$ and $t_6$.
However, the contrast in the fringes survives, since these two symmetric instants are almost entirely responsible 
for the signal at~L2.

Finally, we note that most details in the PMD are preserved between the gaussian and $\sin^2$ envelopes. 
This weak dependence on the envelope function
is undoubtedly highly advantageous when comparing idealized theoretical scenarios with realistic experimental set\-ups,
where the details are most likely not known to high accuracy and controlling them is a serious challenge.

\subsection{Carrier-envelope phase dependence}
  
Next, we analyze the variation of the PMD with the CEP, considering $\phi=0^\circ,45^\circ,90^\circ$, and $135^\circ$, while
leaving all other parameters fixed at the benchmark pulse. The results are displayed in Fig.~\ref{fig:4}, where we show the PMDs
with their corresponding pulse to facilitate the discussion. 

The origin of the shape of the PMD for different CEPs can be understood using similar arguments 
than the ones introduced in Sec.~\ref{sec:3:a} for the specific case of $\phi=90^\circ$. 
Considering first the case $\phi=0^\circ$, one might be tempted to simply interchange the role of the $x$ 
and $y$ axes in the $\phi=90^\circ$ case.  However, the light's ellipticity 
with the major axis along the $y$ direction breaks down this simple picture. Indeed, we clearly 
see in Fig.~\ref{fig:4} that the maxima of the field along the $y$ direction 
still dominate the maximum of the field in the $x$ direction at $t_m$ (the maximum value of the pulse envelope). 
Therefore, the PMD still exhibits its maximum signal at L1 and L2, although
it is understandably weaker than at $\phi=90^\circ$. The fringes at both L1 and L2 now depict high contrast, 
since the signal originates from two nearly equal electron bursts, whereas the contrast at L4 is
now poor, since it is produced mostly by a single electron burst.

As the CEP increases, the peak of the electric field along the positive $y$ axis shifts toward the 
maximum of the pulse envelope, thereby leading to an increase of the signal at L2.  This is accompanied
by a decrease of the contrast of the fringes, as can be seen at $\phi=45^\circ$. As the CEP increases 
to $\phi=90^\circ$, the contrast of the fringes at L3 and L4 deteriorates, since those peaks are mainly produced 
by a single electron burst on the positive and negative $x$ axis. At $\phi=135^\circ$, the contrast 
increases at L4, because it results from an electron emitted at two nearly equal peaks of the field 
along the positive $x$ axis.

It is interesting to look at the difference between the PMDs at $\phi=45^\circ$ and $\phi=135^\circ$, 
i.e., for symmetric situations with respect to $\phi=90^\circ$. 
The main peak of the electric field along the $y$ axis takes the same value in both cases, with the 
difference being that it either occurs after or before $t_m$ for $\phi=45^\circ$ and $\phi=135^\circ$, 
respectively. On the other hand, 
the maximum of the electric field along the $x$ axis switches from the positive ($\phi=45^\circ$) 
to the negative ($\phi=135^\circ$) direction. 
Consequently, the electron is emitted preferentially just before or after the field lies 
along the $y$ axis.  This leads to a noticeable shift, from lower to higher values, of the offset angle at L1 and L2
for $\phi=45^\circ$ and $\phi=135^\circ$, respectively, 

In most attoclock experiments, e.g.~\cite{Satya18}, the CEP is not stabilized.  Consequently, one must average 
the theoretical signal produced by atoms ionized by pulses with randomly distributed CEP to interpret the experimental data.
Even though averaging over the CEP dependence smoothes the results to a nearly spherically symmetric distribution, 
it is still possible to define the principal axes of the ellipse
and extract the offset angle. Pulses with CEPs of $\phi$ and $\phi+180^\circ$ produce PMDs 
that are mirror images of one another by inversion symmetry. 
As a result, the averaged PMD acquires central symmetry. Furthermore, since PMDs with symmetric CEP 
relative to 90$^\circ$ produce either slightly smaller or larger offset angles than would be seen at $\phi=90^\circ$, 
the resulting averaged PMD exhibits an offset angle that should be close to that of the $\phi=90^\circ$ case. 

For narrow $\sin^4$ envelopes, i.e., near-single-cycle pulses, it is relatively straightforward to 
unambiguously extract the offset angle~\cite{Torlina15}. On the other hand, the case of few-cycle pulses with broader envelopes 
is more complicated, as there can exist several maxima at various fringe orders, which actually 
correspond to {\it different\/} offset angles \cite{PhysRevLett.127.273201} depending on the final value of the momentum $p_r=(p_x^2+p_y^2)^{1/2}$.
In fact, it was observed in \cite{PhysRevLett.127.273201} that the offset angle shifts even in a single 
above-threshold ionization (ATI) peak, although the reason could not be identified.
In a realistic experiment, one usually integrates over particular energy bins in order to obtain an average 
offset angle~\cite{Satya18}. 
As a result, it is by no means obvious how to define the atto\-clock angle and, in turn, what information can be
obtained experimentally about the tunneling time even for short pulses. 

The origin of the change in offset angles lies in different exit times and positions in the field, and it is 
intrinsically linked with the action of the Coulomb potential. The 
dependence of the offset angles on the final electron energy, therefore, can provide important physical insights. 
In Sec.~\ref{sec:ellipticity_dependence}, we will show that this effect is enhanced by pulses with 
lower ellipticity and 
provide further explanations for the observed offset angle.
  
\subsection{Intensity dependence}
\label{sec:intensity dependence}

   \begin{figure*}  [t]
 \begin{center}$
  \begin{array}{cc}
  \vspace{-0.47cm}
\includegraphics[width=8.2cm]{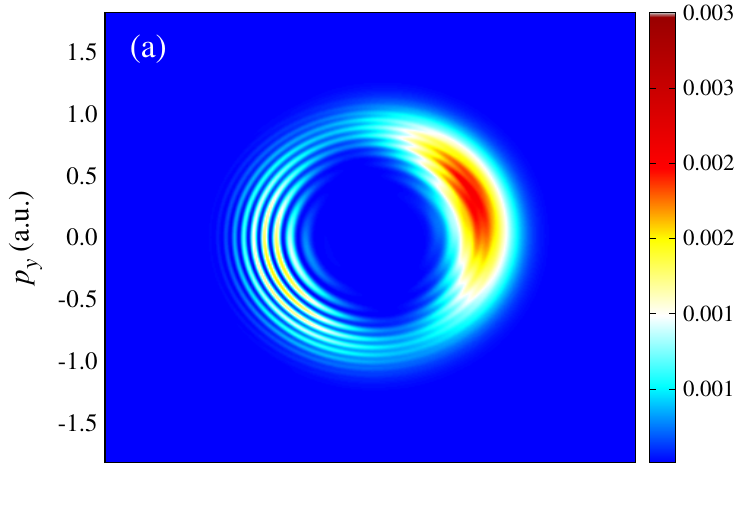} 
&\includegraphics[width=8.2cm]{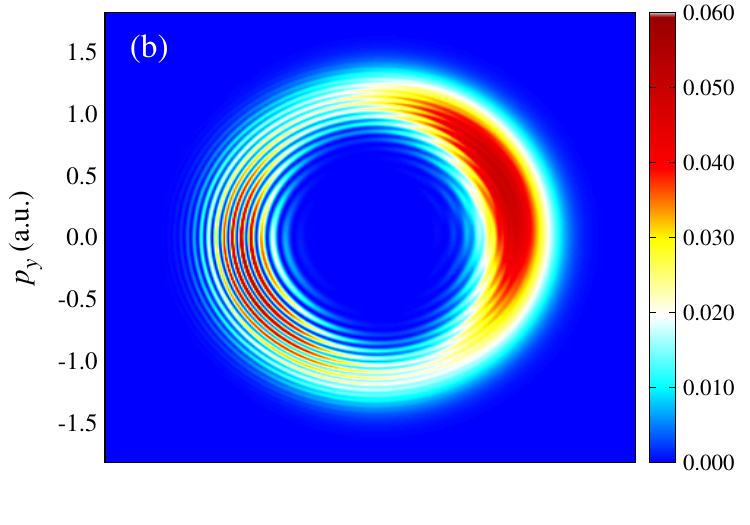} \\
 \includegraphics[width=8.2cm]{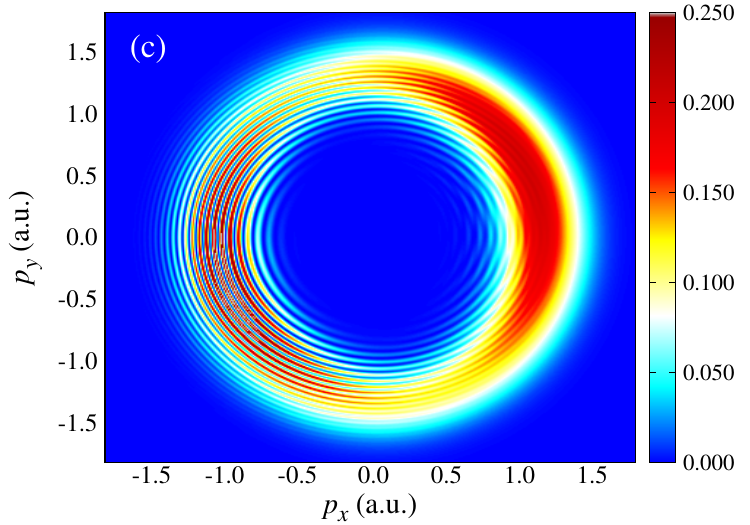} 
&\includegraphics[width=8.2cm]{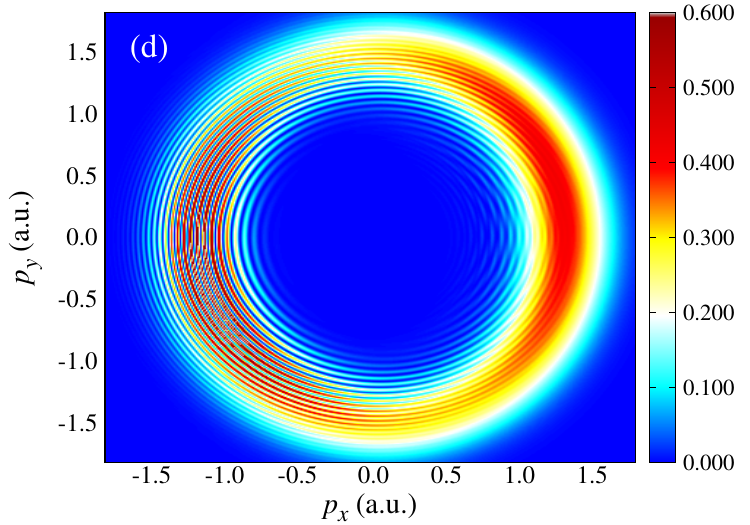}
\end{array}$
   \end{center}
 \caption{PMDs for pulse parameters $N=4$, $\epsilon=0.87$, sin$^2$  envelope, $\phi=90^\circ$, with
\hbox{(a) $I=10^{14}\,$W/cm$^{2}$}, \hbox{(b) $I=2\times 10^{14}\,$W/cm$^{2}$}, \hbox{(c) $I=3\times 10^{14}\,$W/cm$^{2}$}, and 
 \hbox{(d) $I=4\times 10^{14}\,$W/cm$^{2}$}.}  
  \label{fig:5}
  \end{figure*}

Figure~\ref{fig:5} exhibits the dependence of the PMD on the peak intensity of the pulse.
As expected, the ionization signal increases continuously with intensity, thereby indicating that 
we have not yet reached the saturation regime at the highest studied intensity. This fact is not 
surprising considering the short duration of our pulse.
In addition, the mean kinetic energy of the photoelectron becomes larger with increasing pulse intensity, 
which can be seen in a multi\-photon picture as the increasing probability to absorb photons, 
or in the tunneling picture, as the fact that the asymptotic momentum of the electron, neglecting 
the Coulomb potential, is $\bm p=-{\bm A}(t_r)$, where ${\bm A}(t_r)$ is the vector potential at 
the release time $t_r$ of the electron. The latter consideration indicates that the mean asymptotic 
energy of the electron approximately grows linearly with the laser intensity.

Although it is difficult to see it from the figure, the offset angle diminishes slowly with 
intensity \cite{Torlina15,Landsman14,Zimmermann16,PhysRevLett.119.023201,Satya18}. 
This effect can in part be explained by the fact that the action of the Coulomb forces
decreases for fast electrons since they spend less time near the nucleus. The variation of 
the offset angle with intensity is important for realistic
experimental set\-ups, in which not only the CEP is likely not stabilized, but one would also have to
perform focal-averaging to simulate the actual measurements. Due to the broadening of all the features,
the lower peak intensities allow the extraction of the average offset angle with the smallest
uncertainty.  

The variation of the offset angle with intensity can be estimated quantitatively with the Keldysh-Rutherford (KR) model \cite{Bray18}, 
where the offset angle $\theta$ is approximatively given by the Rutherford formula
\begin{eqnarray}
\label{eq:KR}
\tan\theta=\frac{1}{{\rho}}\frac{Z}{v^2_{\infty}}=\frac{1}{{L}}\frac{Z}{v_{\infty}}=\frac{Z}{L}\frac{\omega\sqrt{2}}{E_0}.
\end{eqnarray}
In Eq.~\eqref{eq:KR}, $v_{\infty}=E_0/\sqrt{2}\omega$ is the asymptotic electron speed, where $E_0/\sqrt{2}$ 
is the maximum field strength for circularly polarized light (see Eq.~\eqref{eq:vector_potential}), $\rho$ is 
the impact parameter of the collision, $L=\rho v_{\infty}$ is the asymptotic electron angular momentum of the photoelectron, and
$Z$ is the charge of the residual ion. In the KR model, it is assumed that $\rho\approx r_0$, where $r_0$ is the point 
of closest approach, which is taken as the tunnel exit position at the peak of the electric field, i.e., 
$r_0=I_p\sqrt{2}/E_0$. These assumptions lead to $\tan\theta=\omega^2Z\sqrt{2}/(E_0I_p)$. The latter 
formula can also be found \cite{Bray18} by replacing $L=I_p/\omega$ for a circular pulse in \eqref{eq:KR}. 

The KR formula was employed in \cite{Bray18} to predict, after introducing some fitting parameters, 
the offset angle of different atoms in the circular attoclock.
In particular, the KR model correctly predicts the ${I}^{-1/2}$ dependence of the offset angle on the 
intensity and the change to an ${I}^{-1}$ dependence as one enters the over-the-barrier ionization (OBI) regime \cite{Bray18}. 
Nevertheless, one should keep in mind the limitations of the KR model, as will be discussed further 
in Sec.~\ref{sec:ellipticity_dependence}. While KR can explain the variations of the offset angle at 
the main emission signal, it cannot explain other significant effects visible in the PMD. This is due to the fact that 
KR considers the Coulomb potential to be dominant and does not account for the combined action of the electric 
and Coulomb fields at short range, which becomes utterly important to describe the details of the tunneling 
mechanism in realistic few-cycle pulses used in actual experiments. 
  
Focusing back on Fig.~\ref{fig:5}, we observe that while the overall ionization signal increases, 
the signal could vary differently depending on the regions in the PMD. 
Following the predictions from the Ammosov-Delone-Krainov (ADK) theory, the tunneling ionization rate should increase 
exponentially with field strength \cite{ADK86,Tong02}. This means that the
signal at L1 should become overwhelmingly dominant over the signal produced at other lobes as 
the intensity keeps increasing. 
We do not observe this trend in the figure as the relative intensity between the L1 and L2 lobe decreases. 
Although the signal is rescaled at the different intensities to the maximum signal at L1,
moving from $10^{14}\,$W/cm$^2$ to higher intensities clearly adds signal in the angular regions 
L3 and L4, which are perpendicular to the dominant peak at L1 and 
its weaker counterpart L2. In addition, the signals at L1 and L2 cover a much broader angular range 
as the intensity increases, such that the offset angle becomes difficult to assign. These are clear indications for the onset of the OBI regime, 
which is known to start at a critical intensity $I_c=1.4\times10^{14}\,$W/cm$^2$ and ultimately broadens the angular distribution.

\subsection{Ellipticity dependence}
\label{sec:ellipticity_dependence}

  \begin{figure*}
 \begin{center}$
  \begin{array}{ccc}
  \vspace{-0.47cm}
  \includegraphics[width=5.8cm]{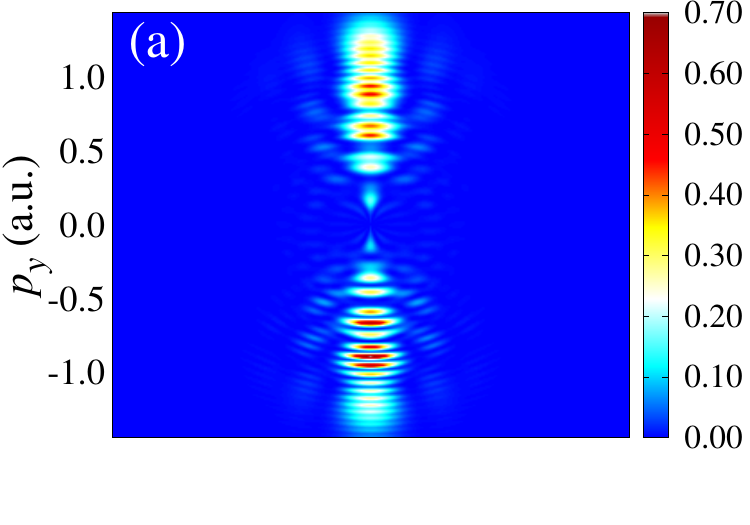} 
  &\includegraphics[width=5.8cm]{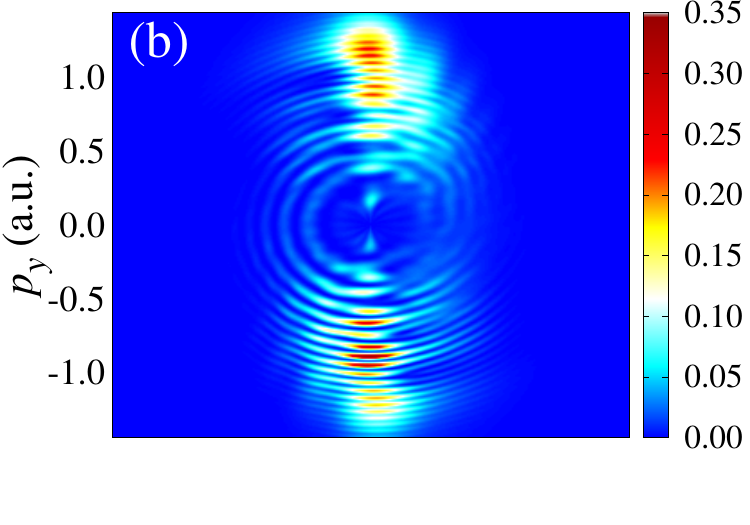} 
  &\includegraphics[width=5.8cm]{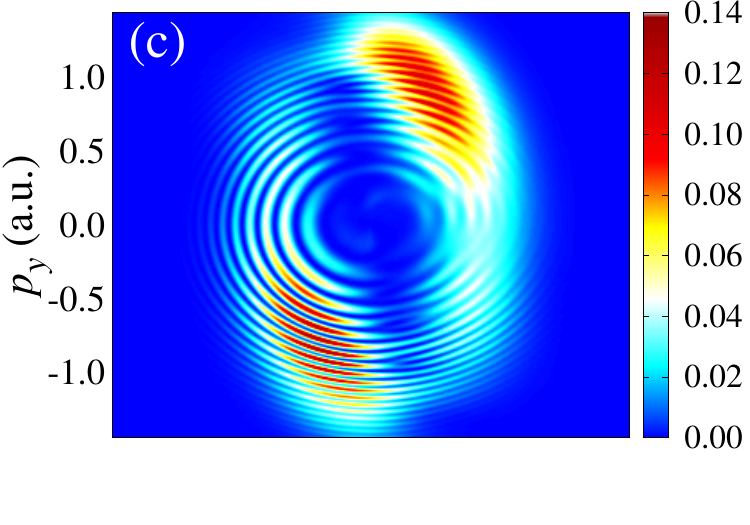}    \vspace{-0.cm}\\
  \includegraphics[width=5.8cm]{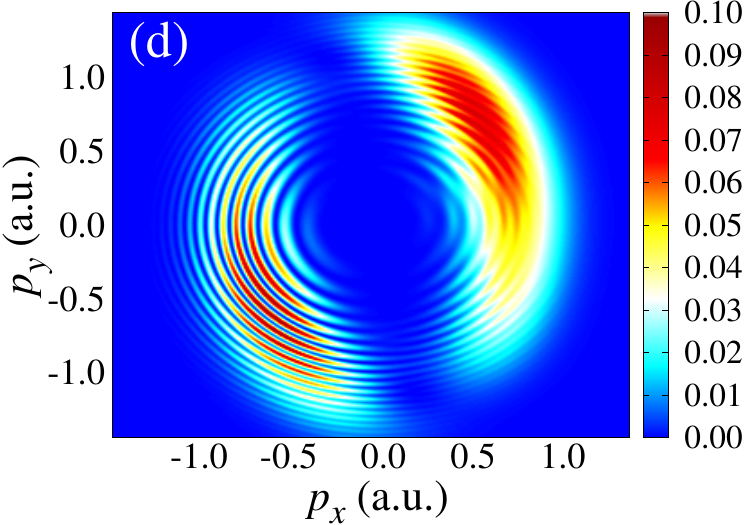}
  &\includegraphics[width=5.8cm]{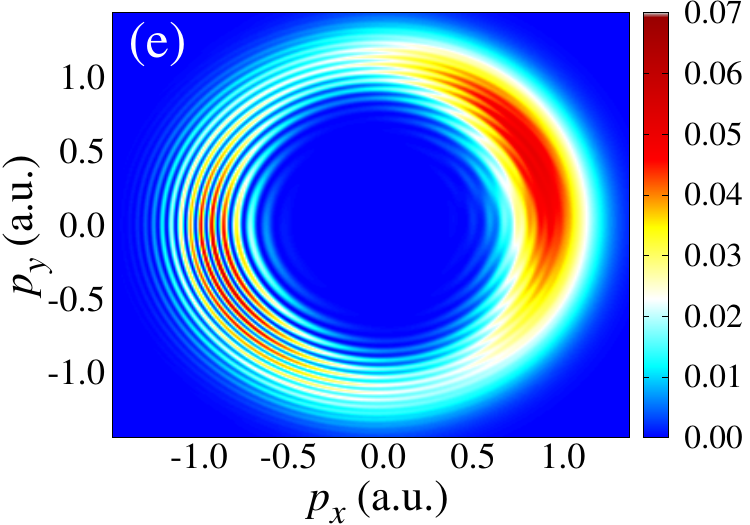}
  &\includegraphics[width=5.8cm]{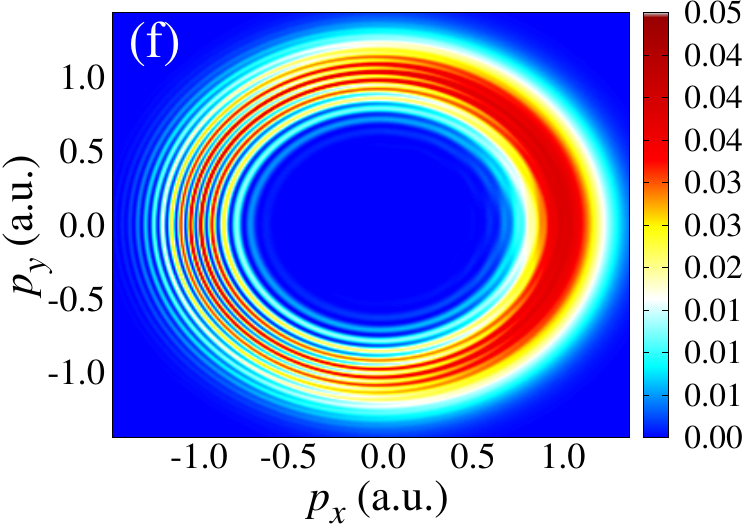}
  \vspace{-0.5cm}
\end{array}$
\end{center}
\caption{PMDs for pulse parameters $N=4$, $I=2\times 10^{14}\,$W/cm$^{2}$, sin$^2$  envelope
 for \hbox{(a) $\epsilon=0.0$}, \hbox{(b) $\epsilon=0.2$}, \hbox{(c) $\epsilon=0.4$}, \hbox{(d) $\epsilon=0.6$},  
 \hbox{(e) $\epsilon=0.87$}, and \hbox{(f) $\epsilon=1.0$}.}  
\label{fig:6}
\end{figure*}

We now turn to the ellipticity dependence of the PMD. Our results are presented in 
Fig.~\ref{fig:6} for pulses with ellipticity $\epsilon=0.0$ (linear)$, 0.2, 0.4, 0.6, 0.87$ and $1.0$ (circular).
The asymptotic momentum  of a classical electron released at the maximum of the electric field, 
and interacting subsequently with the field only, is given by 
${\bm p}=-E_0\epsilon/\omega\sqrt{1+\epsilon^2}\hat{e}_x$. This explains why the maximum signal 
in the PMD increases to higher kinetic energy with increasing light  ellipticity and why the low-energy signal 
almost disappears for nearly circular light. 

The most interesting trend observed in Fig.~\ref{fig:6} concerns the variation of the offset angle 
with ellipticity. This effect was observed experimentally~\cite{Keller12} and theoretically~\cite{PhysRevA.93.023425}
for much longer pulses ($N\ge20$ cycles). The variation of the offset angle with ellipticity at 
the maximum signal can, in part, be explained with a modification of the KR model, specifically by
replacing $v_{\infty}=E_0\epsilon/\omega\sqrt{1+\epsilon^2}$ and $\rho=I_p\sqrt{1+\epsilon^2}/E_0$ in \eqref{eq:KR}.  This leads to 
\begin{eqnarray}
\label{eq:KR-ell}
\tan\theta=\frac{1}{{\rho}}\frac{Z}{v^2_{\infty}}=\frac{\omega^2}{I_p}\frac{\sqrt{1+\epsilon^2}}{E_0\epsilon^2}.
\end{eqnarray}
The above formula correctly predicts that $\theta\to\pi/2$ as $\epsilon\to0$ and that the overall 
offset angle decreases with increasing light ellipticity. A quantitative comparison with the numerical calculations 
is beyond the scope of the present work and would  require a one-cycle pulse with $\sin^4$ envelope and  
fitting parameters, as performed in \cite{Bray18} for the circular case. 

Because the KR model is adapted to situations where the Coulomb effects dominate, i.e., for very short and 
relatively weak pulses, we have confirmed that it is the main origin of the offset angle even for light that 
is not nearly circular by repeating the calculations with a short-range Yukawa potential (as already employed 
in Refs.~\cite{Torlina15,Bray18,Satya18}) to eliminate the Coulomb effects. As expected, we found a zero offset angle even at very low ellipticity. 

Another salient feature visible in the PMDs of the low-ellipticity PMDs in Fig.~\ref{fig:6} is the smooth 
variation of the offset angle with increasing order of the ATI peaks. While the offset angle was also seen 
to increase with kinetic energy in \cite{PhysRevLett.127.273201} for a pulse with $N=20$ cycles and an ellipticity $\epsilon=0.85$, 
the effect is much more pronounced for a 4-cycle pulse with ellipticity $0.4\le\epsilon\le0.6$ as seen in Fig.~\ref{fig:6}. 
Pulses with low ellipticity were also studied in \cite{PhysRevA.93.023425,Keller12}, but the shift of the offset 
angle for different ATI peaks was hardly visible due to the long pulse duration that ultimately blurs the effect. 
Moreover, the energy resolution in \cite{Keller12} was insufficient to observe the individual ATI peaks. 

While various hypotheses were proposed in \cite{PhysRevLett.127.273201} to explain the increase of the offset angle 
with asymptotic electron energy, the main origin of this effect remains unclear. 
In particular, the KR model wrongly predicts that the offset angle at a given ellipticity 
should {\it decrease} with photoelectron energy, since an electron with larger $v_{\infty}$ 
will also have a higher angular momentum $L$ and Eq.~\eqref{eq:KR} clearly indicates a decrease 
of $\theta$ in this case. In fact, the impact parameter $\rho$ (taken as the exit point in KR) 
in \eqref{eq:KR-ell} also increases for a high-energy electron, since it does not tunnel at the 
peak of the electric field. This is another factor contributing to the decrease of the offset angle in KR.

We now suggest an explanation for the increase of the offset 
angle with electron energy and demonstrate that a low-ellipticity PMD 
can act as a selector of the electron release instant. 
Neglecting at first the action of the Coulomb potential, an electron released 
in the electromagnetic field at a time $t_r$ will acquire an asymptotic 
momentum ${\bm p}=-{\bm A}(t_r)$. Because the electron tunnels near the peak 
of the field, with a positive or negative delay, $\Delta t=t_r-t_m$, we obtain, 
to lowest order in $\omega\Delta t\ll 1$, that $p_y/p_x\approx\omega\Delta t/\epsilon$. 
Therefore, an electron tunneling slightly after ($\Delta t\ge0$) or before ($\Delta t\le0$) 
the maximum field strength will have $p_y\ge 0$ or $p_y\le 0$, respectively, while the 
angular spread of the momentum will increase with decreasing ellipticity. In addition, 
for an $N$-cycle pulse with an envelope $f(t)=\sin^{2p}(t)$, the difference
$\Delta K=K-K_0$ between the asymptotic energy $K=\bm p^2/2$ 
of an electron released in the field at $t_m\pm\Delta t$ and the asymptotic energy $K_0$ of an 
electron released at $t_m$ is given, to lowest order, by
\begin{eqnarray}
\Delta K\approx\frac{I\Delta t^2}{1+\epsilon^2}\left(1-\epsilon^2+\frac{p}{4N^2}\epsilon^2\right).
\label{eq:Kinetic}
\end{eqnarray}
Equation \eqref{eq:Kinetic} indicates that the asymptotic energy of a classical electron released 
slightly before or after the peak of the electric field will have a higher asymptotic energy than 
an electron released at the peak of the electric field as long as $\epsilon\le(1+p/4N^2)^{-1/2}$. This effect, which 
is more pronounced at low ellipticity and high intensity, is due to the rapid increase of the 
vector potential along the major axis, which largely compensates the decrease of the smaller 
vector potential in the longitudinal direction. The last term in Eq.~\eqref{eq:Kinetic} is due to 
the pulse envelope and is negligible for long pulses with broad envelopes. Our simulations 
using a short-range Yukawa potential reproduce these predictions: as the light ellipticity decreases, 
the center of the PMD shifts to lower $p_x$ values and the curvature of the signal increases until 
the PMD becomes symmetric with respect to the $p_x=0$ axis for linear light.

\begin{figure*}
 \begin{center}$
  \begin{array}{cc}
  \vspace{-0.47cm}
\includegraphics[width=7.3cm]{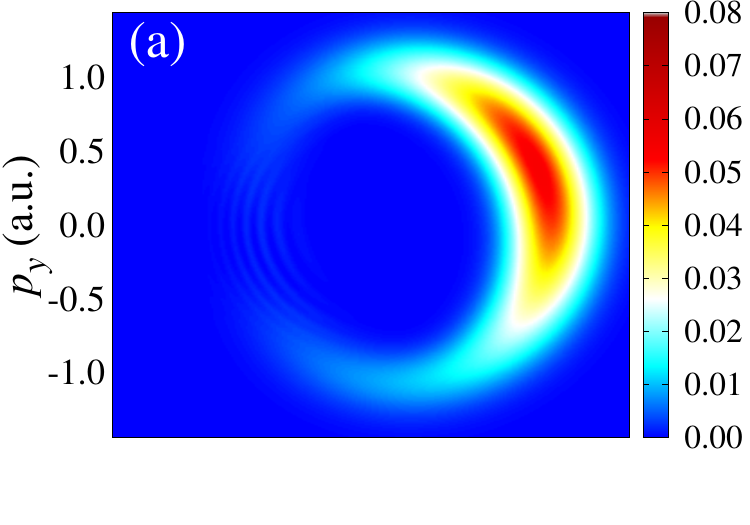} 
&\includegraphics[width=7.3cm]{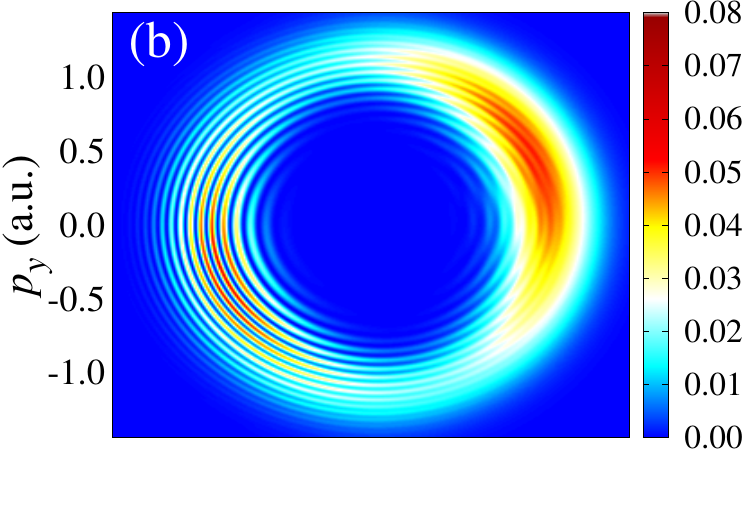} \vspace{-0.2cm}\\
 \includegraphics[width=7.3cm]{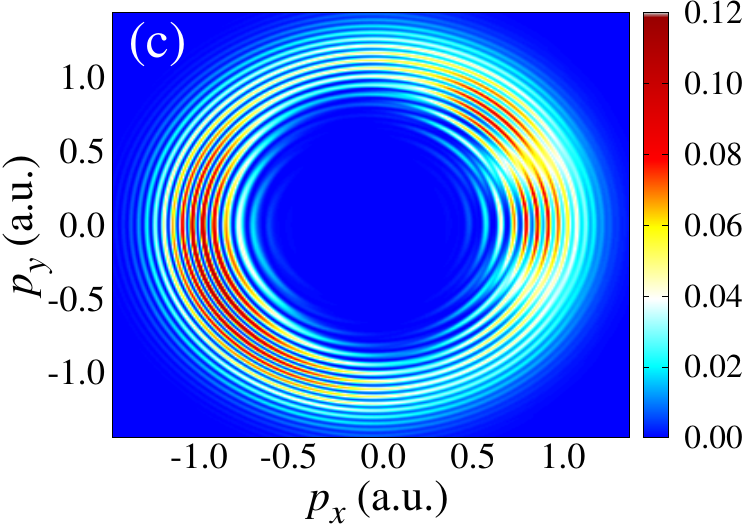} 
&\includegraphics[width=7.3cm]{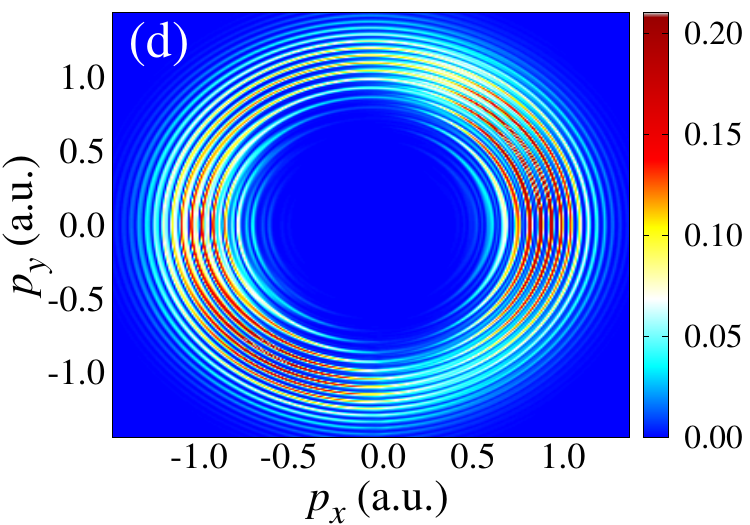}
\vspace{-0.5cm}
\end{array}$
   \end{center}
 \caption{Momentum distributions for pulse parameters $\epsilon=0.87$, $I=2\times 10^{14}\,$W/cm$^{2}$, sin$^2$  envelope,
 $\phi=90^\circ$, for \hbox{(a) $N=2$}, \hbox{(b) $N=4$}, \hbox{(c) $N=6$}, and \hbox{(d) $N=10$}.}  
  \label{fig:7}
  \end{figure*}

Although the introduction of the long-range Coulomb potential breaks this simple picture down, it is still possible 
to draw important conclusions based on the following straightforward considerations. 
Suppose an electron is released at $t_r$ with zero velocity along the electric field at the adiabatic exit position. 
Classical trajectory calculations with a nearly-circular pulse show that, while the offset angle of the 
photo\-electron momentum increases with $\Delta t$ (i.e., an electron released earlier has a smaller offset angle), 
its asymptotic energy remains nearly independent of the release time. Therefore, the main emission signal 
spreads almost equally over a broad angular distribution and 
the energy spread is nearly angle-independent, as can be seen in panels (e) and (f) of Fig.~\ref{fig:6}. 
The situation is different for low-ellipticity pulses, where classical trajectories reveal that both the 
offset angle and the asymptotic electron energy increase with $\Delta t$, such that the main emission 
signal acquires an energy-dependent angular shift that becomes more pronounced with decreasing ellipticity
 (see Fig.~\ref{fig:6}). Thus, a low-ellipticity pulse not only increases the range and the spread of the 
 electron offset angle in the PMD, but it also acts as a selector of the electron release time by 
 dispersing jointly the angular and energy dependence of the signal. 

The origin of the above effect is deceptively simple. For a nearly-circular pulse, the released electron 
is rapidly pulled away by the electric field and does not return near the ionic core. This is a  well-known 
phenomenon  in strong-field physics \cite{PhysRevA.48.R3437,Landsman_2013}. The Coulomb force, therefore, 
only acts effectively when the electron initially escapes the core, such that the electron trajectory gets 
deflected (more strongly for larger $\Delta t$) without significant energy dependence on the release time. For a low-ellipticity pulse, 
the released electron can return closer to the core and the Coulomb force has the possibility to reaccelerate the electron 
upward for a short time, in synchronization with the electric force. Because an electron released at a later time will return 
closer to the core, the action of the Coulomb force is stronger for larger $\Delta t$, resulting in higher 
electron energy. It is important to note that the Coulomb force is the necessary means by which the electron 
experiences an additional upward acceleration. However, it is the electric field that provides the major part of 
the supplementary work. Finally, we emphasize that the above effect can only be understood by taking 
into account the combined action of the Coulomb and electric field.  This explains why a simple model like 
KR cannot grasp its essence.

The above considerations suggest that the study of strong-field physics with short and intense optical 
field with ellipticity $0.4\le \epsilon\le0.6$ could provide important information on the electron 
tunneling dynamics.  This is important for high-harmonic generation as well as the production of 
highly-excited Rydberg states for quantum information.

\subsection{Pulse length dependence}
We finish our analysis with a brief discussion about the dependence of the observed PMD on the length of the pulse.   
As discussed above, the results for the ``near-single cycle'' $N=2$ pulse is straight\-forward to
interpret, with a well-defined maximum at L1 and no observable fringes. Recall, however, that this 
pulse does not have a vanishing
displacement, and hence its physical meaning is questionable. 

While the 4-cycle pulse results in a well-defined 
offset angle, the PMDs for $N=6$ and $N=10$ exhibit complex features that are hard to interpret.  There are just
too many points where the field is sufficiently large to produce substantial tunneling, while the difference
in the field strength at these points is similar enough to fill the entire pattern.  This results in
many fringes, with some additional distortion due to inter\-ference of signals due to multiple emissions at different times. 
As a result, there is no longer a well-defined offset angle, and it is not clear 
whether one can actually extract meaningful tunneling information from the attoclock setup for long pulses.

\section{Conclusions}
\label{sec:4}
In this paper, we have investigated the photo\-electron momentum distribution (PMD) after ionization of 
atomic hydrogen by few-cycle, 800-nm, elliptically polarized light. We analyzed 
and explained the principal features observed in the PMD, including the signal strength, fringe contrast, 
and offset angle in momentum space.  We also analyzed how these features 
are influenced by the pulse characteristics, such as intensity, ellipticity, envelope shape, carrier-envelope phase, and duration.
Evidently, those pulse characteristics reveal important physical insights, 
as the asymptotic electron momentum signal at a specific final kinetic energy 
can be associated with different release times and exit positions in the field. 
Finally, we showed a signature in the PMD for the onset of over-the-barrier ionization.

We hope that the present work will stimulate experimental efforts to further tailor 
few-cycle pulses in order to reveal yet unexplored strong-field effects. 
In particular, we suggest to employ pulses with low ellipticity ($0.4-0.6$) in attoclock set\-ups. 
As shown in this paper, they could become promising tools to unravel tunneling phenomena and information on 
electron wavepackets under the combined effect of the electric field and the Coulomb potential.

\section*{Acknowledgments}
This work~was supported by the United States National 
Science Foundation under grants No.\ \hbox{PHY-1403245} (ND) and No.\ \hbox{PHY-2110023} (KB),
as well as the XSEDE allocation No.\ \hbox{PHY-090031}.


\begin{thebibliography}{42}
\expandafter\ifx\csname natexlab\endcsname\relax\def\natexlab#1{#1}\fi
\expandafter\ifx\csname bibnamefont\endcsname\relax
  \def\bibnamefont#1{#1}\fi
\expandafter\ifx\csname bibfnamefont\endcsname\relax
  \def\bibfnamefont#1{#1}\fi
\expandafter\ifx\csname citenamefont\endcsname\relax
  \def\citenamefont#1{#1}\fi
\expandafter\ifx\csname url\endcsname\relax
  \def\url#1{\texttt{#1}}\fi
\expandafter\ifx\csname urlprefix\endcsname\relax\def\urlprefix{URL }\fi
\providecommand{\bibinfo}[2]{#2}
\providecommand{\eprint}[2][]{\url{#2}}

\bibitem[{\citenamefont{{A. Ferr\'{e}~}{\it et al.}}(2014)}]{Ferre14}
\bibinfo{author}{\bibnamefont{{A. Ferr\'{e}~}{\it et al.}}},
  \bibinfo{journal}{Nat. Photonics} \textbf{\bibinfo{volume}{9}},
  \bibinfo{pages}{93} (\bibinfo{year}{2014}).

\bibitem[{\citenamefont{Neufeld et~al.}(2017)\citenamefont{Neufeld, Bordo,
  Fleischer, and Cohen}}]{photonics4020031}
\bibinfo{author}{\bibfnamefont{O.}~\bibnamefont{Neufeld}},
  \bibinfo{author}{\bibfnamefont{E.}~\bibnamefont{Bordo}},
  \bibinfo{author}{\bibfnamefont{A.}~\bibnamefont{Fleischer}},
  \bibnamefont{and} \bibinfo{author}{\bibfnamefont{O.}~\bibnamefont{Cohen}},
  \bibinfo{journal}{Photonics} \textbf{\bibinfo{volume}{4}}
  (\bibinfo{year}{2017}), ISSN \bibinfo{issn}{2304-6732}.

\bibitem[{\citenamefont{Reich and Madsen}(2016)}]{Reich16}
\bibinfo{author}{\bibfnamefont{D.~M.} \bibnamefont{Reich}} \bibnamefont{and}
  \bibinfo{author}{\bibfnamefont{L.~B.} \bibnamefont{Madsen}},
  \bibinfo{journal}{Phys. Rev. A} \textbf{\bibinfo{volume}{93}},
  \bibinfo{pages}{043411} (\bibinfo{year}{2016}).

\bibitem[{\citenamefont{Mancuso et~al.}(2016)\citenamefont{Mancuso, Dorney,
  Hickstein, Chaloupka, Ellis, Dollar, Knut, Grychtol, Zusin, Gentry
  et~al.}}]{PhysRevLett.117.133201}
\bibinfo{author}{\bibfnamefont{C.~A.} \bibnamefont{Mancuso}},
  \bibinfo{author}{\bibfnamefont{K.~M.} \bibnamefont{Dorney}},
  \bibinfo{author}{\bibfnamefont{D.~D.} \bibnamefont{Hickstein}},
  \bibinfo{author}{\bibfnamefont{J.~L.} \bibnamefont{Chaloupka}},
  \bibinfo{author}{\bibfnamefont{J.~L.} \bibnamefont{Ellis}},
  \bibinfo{author}{\bibfnamefont{F.~J.} \bibnamefont{Dollar}},
  \bibinfo{author}{\bibfnamefont{R.}~\bibnamefont{Knut}},
  \bibinfo{author}{\bibfnamefont{P.}~\bibnamefont{Grychtol}},
  \bibinfo{author}{\bibfnamefont{D.}~\bibnamefont{Zusin}},
  \bibinfo{author}{\bibfnamefont{C.}~\bibnamefont{Gentry}},
  \bibnamefont{et~al.}, \bibinfo{journal}{Phys. Rev. Lett.}
  \textbf{\bibinfo{volume}{117}}, \bibinfo{pages}{133201}
  (\bibinfo{year}{2016}).

\bibitem[{\citenamefont{Long et~al.}(1995)\citenamefont{Long, Becker, and
  McIver}}]{PhysRevA.52.2262}
\bibinfo{author}{\bibfnamefont{S.}~\bibnamefont{Long}},
  \bibinfo{author}{\bibfnamefont{W.}~\bibnamefont{Becker}}, \bibnamefont{and}
  \bibinfo{author}{\bibfnamefont{J.~K.} \bibnamefont{McIver}},
  \bibinfo{journal}{Phys. Rev. A} \textbf{\bibinfo{volume}{52}},
  \bibinfo{pages}{2262} (\bibinfo{year}{1995}).

\bibitem[{\citenamefont{Milo\ifmmode \check{s}\else
  \v{s}\fi{}evi\ifmmode~\acute{c}\else \'{c}\fi{} and
  Becker}(2000)}]{PhysRevA.62.011403}
\bibinfo{author}{\bibfnamefont{D.~B.} \bibnamefont{Milo\ifmmode \check{s}\else
  \v{s}\fi{}evi\ifmmode~\acute{c}\else \'{c}\fi{}}} \bibnamefont{and}
  \bibinfo{author}{\bibfnamefont{W.}~\bibnamefont{Becker}},
  \bibinfo{journal}{Phys. Rev. A} \textbf{\bibinfo{volume}{62}},
  \bibinfo{pages}{011403} (\bibinfo{year}{2000}).

\bibitem[{\citenamefont{Pauly et~al.}(2020)\citenamefont{Pauly, Bondy,
  Hamilton, Douguet, Tong, Chetty, and Bartschat}}]{PhysRevA.102.013116}
\bibinfo{author}{\bibfnamefont{T.}~\bibnamefont{Pauly}},
  \bibinfo{author}{\bibfnamefont{A.}~\bibnamefont{Bondy}},
  \bibinfo{author}{\bibfnamefont{K.~R.} \bibnamefont{Hamilton}},
  \bibinfo{author}{\bibfnamefont{N.}~\bibnamefont{Douguet}},
  \bibinfo{author}{\bibfnamefont{X.-M.} \bibnamefont{Tong}},
  \bibinfo{author}{\bibfnamefont{D.}~\bibnamefont{Chetty}}, \bibnamefont{and}
  \bibinfo{author}{\bibfnamefont{K.}~\bibnamefont{Bartschat}},
  \bibinfo{journal}{Phys. Rev. A} \textbf{\bibinfo{volume}{102}},
  \bibinfo{pages}{013116} (\bibinfo{year}{2020}).

\bibitem[{\citenamefont{De~Silva
  et~al.}(2021{\natexlab{a}})\citenamefont{De~Silva, Atri-Schuller, Dubey,
  Acharya, Romans, Foster, Russ, Compton, Rischbieter, Douguet
  et~al.}}]{PhysRevLett.126.023201}
\bibinfo{author}{\bibfnamefont{A.~H. N.~C.} \bibnamefont{De~Silva}},
  \bibinfo{author}{\bibfnamefont{D.}~\bibnamefont{Atri-Schuller}},
  \bibinfo{author}{\bibfnamefont{S.}~\bibnamefont{Dubey}},
  \bibinfo{author}{\bibfnamefont{B.~P.} \bibnamefont{Acharya}},
  \bibinfo{author}{\bibfnamefont{K.~L.} \bibnamefont{Romans}},
  \bibinfo{author}{\bibfnamefont{K.}~\bibnamefont{Foster}},
  \bibinfo{author}{\bibfnamefont{O.}~\bibnamefont{Russ}},
  \bibinfo{author}{\bibfnamefont{K.}~\bibnamefont{Compton}},
  \bibinfo{author}{\bibfnamefont{C.}~\bibnamefont{Rischbieter}},
  \bibinfo{author}{\bibfnamefont{N.}~\bibnamefont{Douguet}},
  \bibnamefont{et~al.}, \bibinfo{journal}{Phys. Rev. Lett.}
  \textbf{\bibinfo{volume}{126}}, \bibinfo{pages}{023201}
  (\bibinfo{year}{2021}{\natexlab{a}}).

\bibitem[{\citenamefont{De~Silva
  et~al.}(2021{\natexlab{b}})\citenamefont{De~Silva, Moon, Romans, Acharya,
  Dubey, Foster, Russ, Rischbieter, Douguet, Bartschat
  et~al.}}]{PhysRevA.103.053125}
\bibinfo{author}{\bibfnamefont{A.~H. N.~C.} \bibnamefont{De~Silva}},
  \bibinfo{author}{\bibfnamefont{T.}~\bibnamefont{Moon}},
  \bibinfo{author}{\bibfnamefont{K.~L.} \bibnamefont{Romans}},
  \bibinfo{author}{\bibfnamefont{B.~P.} \bibnamefont{Acharya}},
  \bibinfo{author}{\bibfnamefont{S.}~\bibnamefont{Dubey}},
  \bibinfo{author}{\bibfnamefont{K.}~\bibnamefont{Foster}},
  \bibinfo{author}{\bibfnamefont{O.}~\bibnamefont{Russ}},
  \bibinfo{author}{\bibfnamefont{C.}~\bibnamefont{Rischbieter}},
  \bibinfo{author}{\bibfnamefont{N.}~\bibnamefont{Douguet}},
  \bibinfo{author}{\bibfnamefont{K.}~\bibnamefont{Bartschat}},
  \bibnamefont{et~al.}, \bibinfo{journal}{Phys. Rev. A}
  \textbf{\bibinfo{volume}{103}}, \bibinfo{pages}{053125}
  (\bibinfo{year}{2021}{\natexlab{b}}).

\bibitem[{\citenamefont{Torlina et~al.}(2015)\citenamefont{Torlina, Morales,
  Kaushal, Ivanov, Kheifets, Zielinski, Scrinzi, Muller, Sukiasyan, Ivanov
  et~al.}}]{Torlina15}
\bibinfo{author}{\bibfnamefont{L.}~\bibnamefont{Torlina}},
  \bibinfo{author}{\bibfnamefont{F.}~\bibnamefont{Morales}},
  \bibinfo{author}{\bibfnamefont{J.}~\bibnamefont{Kaushal}},
  \bibinfo{author}{\bibfnamefont{I.}~\bibnamefont{Ivanov}},
  \bibinfo{author}{\bibfnamefont{A.}~\bibnamefont{Kheifets}},
  \bibinfo{author}{\bibfnamefont{A.}~\bibnamefont{Zielinski}},
  \bibinfo{author}{\bibfnamefont{A.}~\bibnamefont{Scrinzi}},
  \bibinfo{author}{\bibfnamefont{H.~G.} \bibnamefont{Muller}},
  \bibinfo{author}{\bibfnamefont{S.}~\bibnamefont{Sukiasyan}},
  \bibinfo{author}{\bibfnamefont{M.}~\bibnamefont{Ivanov}},
  \bibnamefont{et~al.}, \bibinfo{journal}{Nat. Phys.}
  \textbf{\bibinfo{volume}{11}}, \bibinfo{pages}{503} (\bibinfo{year}{2015}).

\bibitem[{\citenamefont{Landsman et~al.}(2014)\citenamefont{Landsman, Weger,
  Maurer, Boge, Ludwig, Heuser, Cirelli, Gallmann, and Keller}}]{Landsman14}
\bibinfo{author}{\bibfnamefont{A.~S.} \bibnamefont{Landsman}},
  \bibinfo{author}{\bibfnamefont{M.}~\bibnamefont{Weger}},
  \bibinfo{author}{\bibfnamefont{J.}~\bibnamefont{Maurer}},
  \bibinfo{author}{\bibfnamefont{R.}~\bibnamefont{Boge}},
  \bibinfo{author}{\bibfnamefont{A.}~\bibnamefont{Ludwig}},
  \bibinfo{author}{\bibfnamefont{S.}~\bibnamefont{Heuser}},
  \bibinfo{author}{\bibfnamefont{C.}~\bibnamefont{Cirelli}},
  \bibinfo{author}{\bibfnamefont{L.}~\bibnamefont{Gallmann}}, \bibnamefont{and}
  \bibinfo{author}{\bibfnamefont{U.}~\bibnamefont{Keller}},
  \bibinfo{journal}{Optica} \textbf{\bibinfo{volume}{1}}, \bibinfo{pages}{343}
  (\bibinfo{year}{2014}).

\bibitem[{\citenamefont{{T. Zimmermann} et~al.}(2016)\citenamefont{{T.
  Zimmermann}, Mishra, Doran, Gordon, and Landsman}}]{Zimmermann16}
\bibinfo{author}{\bibnamefont{{T. Zimmermann}}},
  \bibinfo{author}{\bibfnamefont{S.}~\bibnamefont{Mishra}},
  \bibinfo{author}{\bibfnamefont{B.~R.} \bibnamefont{Doran}},
  \bibinfo{author}{\bibfnamefont{D.~F.} \bibnamefont{Gordon}},
  \bibnamefont{and} \bibinfo{author}{\bibfnamefont{A.~S.}
  \bibnamefont{Landsman}}, \bibinfo{journal}{Phys. Rev. Lett.}
  \textbf{\bibinfo{volume}{116}}, \bibinfo{pages}{233603}
  (\bibinfo{year}{2016}).

\bibitem[{\citenamefont{Camus et~al.}(2017)\citenamefont{Camus, Yakaboylu,
  Fechner, Klaiber, Laux, Mi, Hatsagortsyan, Pfeifer, Keitel, and
  Moshammer}}]{PhysRevLett.119.023201}
\bibinfo{author}{\bibfnamefont{N.}~\bibnamefont{Camus}},
  \bibinfo{author}{\bibfnamefont{E.}~\bibnamefont{Yakaboylu}},
  \bibinfo{author}{\bibfnamefont{L.}~\bibnamefont{Fechner}},
  \bibinfo{author}{\bibfnamefont{M.}~\bibnamefont{Klaiber}},
  \bibinfo{author}{\bibfnamefont{M.}~\bibnamefont{Laux}},
  \bibinfo{author}{\bibfnamefont{Y.}~\bibnamefont{Mi}},
  \bibinfo{author}{\bibfnamefont{K.~Z.} \bibnamefont{Hatsagortsyan}},
  \bibinfo{author}{\bibfnamefont{T.}~\bibnamefont{Pfeifer}},
  \bibinfo{author}{\bibfnamefont{C.~H.} \bibnamefont{Keitel}},
  \bibnamefont{and}
  \bibinfo{author}{\bibfnamefont{R.}~\bibnamefont{Moshammer}},
  \bibinfo{journal}{Phys. Rev. Lett.} \textbf{\bibinfo{volume}{119}},
  \bibinfo{pages}{023201} (\bibinfo{year}{2017}).

\bibitem[{\citenamefont{Sainadh et~al.}(2019)\citenamefont{Sainadh, Xu, Wang,
  Atia-Tul-Noor, Wallace, Douguet, Bray, Ivanov, Bartschat, Kheifets
  et~al.}}]{Satya18}
\bibinfo{author}{\bibfnamefont{U.~S.} \bibnamefont{Sainadh}},
  \bibinfo{author}{\bibfnamefont{H.}~\bibnamefont{Xu}},
  \bibinfo{author}{\bibfnamefont{X.}~\bibnamefont{Wang}},
  \bibinfo{author}{\bibnamefont{Atia-Tul-Noor}},
  \bibinfo{author}{\bibfnamefont{W.~C.} \bibnamefont{Wallace}},
  \bibinfo{author}{\bibfnamefont{N.}~\bibnamefont{Douguet}},
  \bibinfo{author}{\bibfnamefont{A.~W.} \bibnamefont{Bray}},
  \bibinfo{author}{\bibfnamefont{I.}~\bibnamefont{Ivanov}},
  \bibinfo{author}{\bibfnamefont{K.}~\bibnamefont{Bartschat}},
  \bibinfo{author}{\bibfnamefont{A.}~\bibnamefont{Kheifets}},
  \bibnamefont{et~al.}, \bibinfo{journal}{Nature}
  \textbf{\bibinfo{volume}{75}}, \bibinfo{pages}{75} (\bibinfo{year}{2019}).

\bibitem[{\citenamefont{Han et~al.}(2018)\citenamefont{Han, Ge, Shao, Gong, and
  Liu}}]{Peipei18}
\bibinfo{author}{\bibfnamefont{M.}~\bibnamefont{Han}},
  \bibinfo{author}{\bibfnamefont{P.}~\bibnamefont{Ge}},
  \bibinfo{author}{\bibfnamefont{Y.}~\bibnamefont{Shao}},
  \bibinfo{author}{\bibfnamefont{Q.}~\bibnamefont{Gong}}, \bibnamefont{and}
  \bibinfo{author}{\bibfnamefont{Y.}~\bibnamefont{Liu}},
  \bibinfo{journal}{Phys. Rev. Lett.} \textbf{\bibinfo{volume}{120}},
  \bibinfo{pages}{073202} (\bibinfo{year}{2018}).

\bibitem[{\citenamefont{Ge et~al.}(2019)\citenamefont{Ge, Han, Deng, Gong, and
  Liu}}]{Peipei19}
\bibinfo{author}{\bibfnamefont{P.}~\bibnamefont{Ge}},
  \bibinfo{author}{\bibfnamefont{M.}~\bibnamefont{Han}},
  \bibinfo{author}{\bibfnamefont{Y.}~\bibnamefont{Deng}},
  \bibinfo{author}{\bibfnamefont{Q.}~\bibnamefont{Gong}}, \bibnamefont{and}
  \bibinfo{author}{\bibfnamefont{Y.}~\bibnamefont{Liu}},
  \bibinfo{journal}{Phys. Rev. Lett.} \textbf{\bibinfo{volume}{122}},
  \bibinfo{pages}{013201} (\bibinfo{year}{2019}).

\bibitem[{\citenamefont{Trabert
  et~al.}(2021{\natexlab{a}})\citenamefont{Trabert, Anders, Brennecke,
  Sch\"offler, Jahnke, Schmidt, Kunitski, Lein, D\"orner, and
  Eckart}}]{PhysRevLett.127.273201}
\bibinfo{author}{\bibfnamefont{D.}~\bibnamefont{Trabert}},
  \bibinfo{author}{\bibfnamefont{N.}~\bibnamefont{Anders}},
  \bibinfo{author}{\bibfnamefont{S.}~\bibnamefont{Brennecke}},
  \bibinfo{author}{\bibfnamefont{M.~S.} \bibnamefont{Sch\"offler}},
  \bibinfo{author}{\bibfnamefont{T.}~\bibnamefont{Jahnke}},
  \bibinfo{author}{\bibfnamefont{L.~P.~H.} \bibnamefont{Schmidt}},
  \bibinfo{author}{\bibfnamefont{M.}~\bibnamefont{Kunitski}},
  \bibinfo{author}{\bibfnamefont{M.}~\bibnamefont{Lein}},
  \bibinfo{author}{\bibfnamefont{R.}~\bibnamefont{D\"orner}}, \bibnamefont{and}
  \bibinfo{author}{\bibfnamefont{S.}~\bibnamefont{Eckart}},
  \bibinfo{journal}{Phys. Rev. Lett.} \textbf{\bibinfo{volume}{127}},
  \bibinfo{pages}{273201} (\bibinfo{year}{2021}{\natexlab{a}}).

\bibitem[{\citenamefont{Trabert
  et~al.}(2021{\natexlab{b}})\citenamefont{Trabert, Brennecke, Fehre, Anders,
  Geyer, Grundmann, Schöffler, Schmidt, Jahnke, Dörner et~al.}}]{Trabert21}
\bibinfo{author}{\bibfnamefont{D.}~\bibnamefont{Trabert}},
  \bibinfo{author}{\bibfnamefont{S.}~\bibnamefont{Brennecke}},
  \bibinfo{author}{\bibfnamefont{K.}~\bibnamefont{Fehre}},
  \bibinfo{author}{\bibfnamefont{N.}~\bibnamefont{Anders}},
  \bibinfo{author}{\bibfnamefont{A.}~\bibnamefont{Geyer}},
  \bibinfo{author}{\bibfnamefont{S.}~\bibnamefont{Grundmann}},
  \bibinfo{author}{\bibfnamefont{M.~S.} \bibnamefont{Schöffler}},
  \bibinfo{author}{\bibfnamefont{L.~P.~H.} \bibnamefont{Schmidt}},
  \bibinfo{author}{\bibfnamefont{T.}~\bibnamefont{Jahnke}},
  \bibinfo{author}{\bibfnamefont{R.}~\bibnamefont{Dörner}},
  \bibnamefont{et~al.}, \bibinfo{journal}{Nat. Comm.}
  \textbf{\bibinfo{volume}{12}}, \bibinfo{pages}{1697}
  (\bibinfo{year}{2021}{\natexlab{b}}).

\bibitem[{\citenamefont{Wang et~al.}(2014)\citenamefont{Wang, Lai, Hu, Chen,
  Quan, Kang, Gong, and Liu}}]{Wang14}
\bibinfo{author}{\bibfnamefont{C.}~\bibnamefont{Wang}},
  \bibinfo{author}{\bibfnamefont{X.}~\bibnamefont{Lai}},
  \bibinfo{author}{\bibfnamefont{Z.}~\bibnamefont{Hu}},
  \bibinfo{author}{\bibfnamefont{Y.}~\bibnamefont{Chen}},
  \bibinfo{author}{\bibfnamefont{W.}~\bibnamefont{Quan}},
  \bibinfo{author}{\bibfnamefont{H.}~\bibnamefont{Kang}},
  \bibinfo{author}{\bibfnamefont{C.}~\bibnamefont{Gong}}, \bibnamefont{and}
  \bibinfo{author}{\bibfnamefont{X.}~\bibnamefont{Liu}},
  \bibinfo{journal}{Phys. Rev. A} \textbf{\bibinfo{volume}{90}},
  \bibinfo{pages}{013422} (\bibinfo{year}{2014}).

\bibitem[{\citenamefont{Eckle et~al.}(2008)\citenamefont{Eckle, Pfeiffer,
  Cirelli, Staudte, D{\"o}rner, Muller, B{\"u}ttiker, and Keller}}]{Eckle1525}
\bibinfo{author}{\bibfnamefont{P.}~\bibnamefont{Eckle}},
  \bibinfo{author}{\bibfnamefont{A.~N.} \bibnamefont{Pfeiffer}},
  \bibinfo{author}{\bibfnamefont{C.}~\bibnamefont{Cirelli}},
  \bibinfo{author}{\bibfnamefont{A.}~\bibnamefont{Staudte}},
  \bibinfo{author}{\bibfnamefont{R.}~\bibnamefont{D{\"o}rner}},
  \bibinfo{author}{\bibfnamefont{H.~G.} \bibnamefont{Muller}},
  \bibinfo{author}{\bibfnamefont{M.}~\bibnamefont{B{\"u}ttiker}},
  \bibnamefont{and} \bibinfo{author}{\bibfnamefont{U.}~\bibnamefont{Keller}},
  \bibinfo{journal}{Science} \textbf{\bibinfo{volume}{322}},
  \bibinfo{pages}{1525} (\bibinfo{year}{2008}).

\bibitem[{\citenamefont{Pfeiffer
  et~al.}(2012{\natexlab{a}})\citenamefont{Pfeiffer, Cirelli, Smolarski,
  Dimitrovski, Abusamha, Madsen, and Keller}}]{Pfeiffer12}
\bibinfo{author}{\bibfnamefont{A.~N.} \bibnamefont{Pfeiffer}},
  \bibinfo{author}{\bibfnamefont{C.}~\bibnamefont{Cirelli}},
  \bibinfo{author}{\bibfnamefont{M.}~\bibnamefont{Smolarski}},
  \bibinfo{author}{\bibfnamefont{D.}~\bibnamefont{Dimitrovski}},
  \bibinfo{author}{\bibfnamefont{M.}~\bibnamefont{Abusamha}},
  \bibinfo{author}{\bibfnamefont{L.~B.} \bibnamefont{Madsen}},
  \bibnamefont{and} \bibinfo{author}{\bibfnamefont{U.}~\bibnamefont{Keller}},
  \bibinfo{journal}{Nat. Phys.} \textbf{\bibinfo{volume}{8}},
  \bibinfo{pages}{76} (\bibinfo{year}{2012}{\natexlab{a}}).

\bibitem[{\citenamefont{Hanus et~al.}(2018)\citenamefont{Hanus, Kangaparambil,
  Larimian, Xie, Sch\"{o}ffler, Staudte, Paulus, Baltuska, and
  Kitzler}}]{Hanus18}
\bibinfo{author}{\bibfnamefont{V.}~\bibnamefont{Hanus}},
  \bibinfo{author}{\bibfnamefont{S.}~\bibnamefont{Kangaparambil}},
  \bibinfo{author}{\bibfnamefont{S.}~\bibnamefont{Larimian}},
  \bibinfo{author}{\bibfnamefont{X.}~\bibnamefont{Xie}},
  \bibinfo{author}{\bibfnamefont{M.}~\bibnamefont{Sch\"{o}ffler}},
  \bibinfo{author}{\bibfnamefont{A.}~\bibnamefont{Staudte}},
  \bibinfo{author}{\bibfnamefont{G.}~\bibnamefont{Paulus}},
  \bibinfo{author}{\bibfnamefont{A.}~\bibnamefont{Baltuska}}, \bibnamefont{and}
  \bibinfo{author}{\bibfnamefont{M.}~\bibnamefont{Kitzler}},
  \bibinfo{journal}{High-Brightness Sources and Light-Driven Interactions} p.
  \bibinfo{pages}{HM4A.4} (\bibinfo{year}{2018}).

\bibitem[{\citenamefont{Serov et~al.}(2019)\citenamefont{Serov, Bray, and
  Kheifets}}]{Serov19}
\bibinfo{author}{\bibfnamefont{V.~V.} \bibnamefont{Serov}},
  \bibinfo{author}{\bibfnamefont{A.~W.} \bibnamefont{Bray}}, \bibnamefont{and}
  \bibinfo{author}{\bibfnamefont{A.~S.} \bibnamefont{Kheifets}},
  \bibinfo{journal}{Phys. Rev. A} \textbf{\bibinfo{volume}{99}},
  \bibinfo{pages}{063428} (\bibinfo{year}{2019}).

\bibitem[{\citenamefont{Sainadh et~al.}(2020)\citenamefont{Sainadh, Sang, and
  Litvinyuk}}]{Satya2020}
\bibinfo{author}{\bibfnamefont{U.~S.} \bibnamefont{Sainadh}},
  \bibinfo{author}{\bibfnamefont{R.~T.} \bibnamefont{Sang}}, \bibnamefont{and}
  \bibinfo{author}{\bibfnamefont{I.~V.} \bibnamefont{Litvinyuk}},
  \bibinfo{journal}{J.~Phys.: Photonics} \textbf{\bibinfo{volume}{2}},
  \bibinfo{pages}{042002} (\bibinfo{year}{2020}).

\bibitem[{\citenamefont{Kheifets}(2020)}]{Kheifets2020}
\bibinfo{author}{\bibfnamefont{A.~S.} \bibnamefont{Kheifets}},
  \bibinfo{journal}{J.~Phys.~B: At.\ Mol.\ Opt. Phys.}
  \textbf{\bibinfo{volume}{53}}, \bibinfo{pages}{072001}
  (\bibinfo{year}{2020}).

\bibitem[{\citenamefont{Douguet and Bartschat}(2019)}]{Douguet19}
\bibinfo{author}{\bibfnamefont{N.}~\bibnamefont{Douguet}} \bibnamefont{and}
  \bibinfo{author}{\bibfnamefont{K.}~\bibnamefont{Bartschat}},
  \bibinfo{journal}{Phys. Rev. A} \textbf{\bibinfo{volume}{99}},
  \bibinfo{pages}{023417} (\bibinfo{year}{2019}).

\bibitem[{\citenamefont{Armstrong et~al.}(2020)\citenamefont{Armstrong, Clarke,
  Benda, Brown, and van~der Hart}}]{Armstrong20}
\bibinfo{author}{\bibfnamefont{G.~S.~J.} \bibnamefont{Armstrong}},
  \bibinfo{author}{\bibfnamefont{D.~D.~A.} \bibnamefont{Clarke}},
  \bibinfo{author}{\bibfnamefont{J.}~\bibnamefont{Benda}},
  \bibinfo{author}{\bibfnamefont{A.~C.} \bibnamefont{Brown}}, \bibnamefont{and}
  \bibinfo{author}{\bibfnamefont{H.~W.} \bibnamefont{van~der Hart}},
  \bibinfo{journal}{Phys. Rev. A} \textbf{\bibinfo{volume}{101}},
  \bibinfo{pages}{041401} (\bibinfo{year}{2020}).

\bibitem[{\citenamefont{Bartschat et~al.}(2015)\citenamefont{Bartschat, Venzke,
  and Grum-Grzhimailo}}]{PhysRevA.91.053404}
\bibinfo{author}{\bibfnamefont{K.}~\bibnamefont{Bartschat}},
  \bibinfo{author}{\bibfnamefont{J.}~\bibnamefont{Venzke}}, \bibnamefont{and}
  \bibinfo{author}{\bibfnamefont{A.~N.} \bibnamefont{Grum-Grzhimailo}},
  \bibinfo{journal}{Phys. Rev. A} \textbf{\bibinfo{volume}{91}},
  \bibinfo{pages}{053404} (\bibinfo{year}{2015}).

\bibitem[{\citenamefont{Li and Jones}(2014)}]{PhysRevLett.112.143006}
\bibinfo{author}{\bibfnamefont{S.}~\bibnamefont{Li}} \bibnamefont{and}
  \bibinfo{author}{\bibfnamefont{R.~R.} \bibnamefont{Jones}},
  \bibinfo{journal}{Phys. Rev. Lett.} \textbf{\bibinfo{volume}{112}},
  \bibinfo{pages}{143006} (\bibinfo{year}{2014}).

\bibitem[{\citenamefont{Joachain et~al.}(2011)\citenamefont{Joachain, Kylstra,
  and Potvliege}}]{Joachain2011}
\bibinfo{author}{\bibfnamefont{C.~J.} \bibnamefont{Joachain}},
  \bibinfo{author}{\bibfnamefont{N.~J.} \bibnamefont{Kylstra}},
  \bibnamefont{and} \bibinfo{author}{\bibfnamefont{R.~M.}
  \bibnamefont{Potvliege}}, \emph{\bibinfo{title}{Atoms in Intense Laser
  Fields}} (\bibinfo{publisher}{Cambridge University Press},
  \bibinfo{year}{2011}).

\bibitem[{\citenamefont{Ivanov et~al.}(2014)\citenamefont{Ivanov, Kheifets,
  Bartschat, Emmons, Buczek, Gryzlova, and
  Grum-Grzhimailo}}]{PhysRevA.90.043401}
\bibinfo{author}{\bibfnamefont{I.~A.} \bibnamefont{Ivanov}},
  \bibinfo{author}{\bibfnamefont{A.~S.} \bibnamefont{Kheifets}},
  \bibinfo{author}{\bibfnamefont{K.}~\bibnamefont{Bartschat}},
  \bibinfo{author}{\bibfnamefont{J.}~\bibnamefont{Emmons}},
  \bibinfo{author}{\bibfnamefont{S.~M.} \bibnamefont{Buczek}},
  \bibinfo{author}{\bibfnamefont{E.~V.} \bibnamefont{Gryzlova}},
  \bibnamefont{and} \bibinfo{author}{\bibfnamefont{A.~N.}
  \bibnamefont{Grum-Grzhimailo}}, \bibinfo{journal}{Phys. Rev. A}
  \textbf{\bibinfo{volume}{90}}, \bibinfo{pages}{043401}
  (\bibinfo{year}{2014}).

\bibitem[{\citenamefont{Xiao et~al.}(2019)\citenamefont{Xiao, Wang, Liang,
  Gong, and Peng}}]{Xiao19}
\bibinfo{author}{\bibfnamefont{X.-R.} \bibnamefont{Xiao}},
  \bibinfo{author}{\bibfnamefont{M.-X.} \bibnamefont{Wang}},
  \bibinfo{author}{\bibfnamefont{H.}~\bibnamefont{Liang}},
  \bibinfo{author}{\bibfnamefont{Q.}~\bibnamefont{Gong}}, \bibnamefont{and}
  \bibinfo{author}{\bibfnamefont{L.-Y.} \bibnamefont{Peng}},
  \bibinfo{journal}{Phys. Rev. Lett.} \textbf{\bibinfo{volume}{122}},
  \bibinfo{pages}{053201} (\bibinfo{year}{2019}).

\bibitem[{\citenamefont{Murakami and Chu}(2016)}]{PhysRevA.93.023425}
\bibinfo{author}{\bibfnamefont{M.}~\bibnamefont{Murakami}} \bibnamefont{and}
  \bibinfo{author}{\bibfnamefont{S.-I.} \bibnamefont{Chu}},
  \bibinfo{journal}{Phys. Rev. A} \textbf{\bibinfo{volume}{93}},
  \bibinfo{pages}{023425} (\bibinfo{year}{2016}).

\bibitem[{\citenamefont{Cormier and Lambropoulos}(1996)}]{CorLam1996}
\bibinfo{author}{\bibfnamefont{E.}~\bibnamefont{Cormier}} \bibnamefont{and}
  \bibinfo{author}{\bibfnamefont{P.}~\bibnamefont{Lambropoulos}},
  \bibinfo{journal}{J. Phys. B} \textbf{\bibinfo{volume}{29}},
  \bibinfo{pages}{1667} (\bibinfo{year}{1996}).

\bibitem[{\citenamefont{Grum-Grzhimailo
  et~al.}(2010)\citenamefont{Grum-Grzhimailo, Abeln, Bartschat, \hbox{Weflen},
  and Urness}}]{PhysRevA.81.043408}
\bibinfo{author}{\bibfnamefont{A.~N.} \bibnamefont{Grum-Grzhimailo}},
  \bibinfo{author}{\bibfnamefont{B.}~\bibnamefont{Abeln}},
  \bibinfo{author}{\bibfnamefont{K.}~\bibnamefont{Bartschat}},
  \bibinfo{author}{\bibfnamefont{D.}~\bibnamefont{\hbox{Weflen}}},
  \bibnamefont{and} \bibinfo{author}{\bibfnamefont{T.}~\bibnamefont{Urness}},
  \bibinfo{journal}{Phys. Rev. A} \textbf{\bibinfo{volume}{81}},
  \bibinfo{pages}{043408} (\bibinfo{year}{2010}).

\bibitem[{\citenamefont{Bray et~al.}(2018)\citenamefont{Bray, Eckart, and
  Kheifets}}]{Bray18}
\bibinfo{author}{\bibfnamefont{A.~W.} \bibnamefont{Bray}},
  \bibinfo{author}{\bibfnamefont{S.}~\bibnamefont{Eckart}}, \bibnamefont{and}
  \bibinfo{author}{\bibfnamefont{A.~S.} \bibnamefont{Kheifets}},
  \bibinfo{journal}{Phys. Rev. Lett.} \textbf{\bibinfo{volume}{121}},
  \bibinfo{pages}{123201} (\bibinfo{year}{2018}).

\bibitem[{\citenamefont{Ammosov et~al.}(1986)\citenamefont{Ammosov, Delone, and
  Krainov}}]{ADK86}
\bibinfo{author}{\bibfnamefont{M.}~\bibnamefont{Ammosov}},
  \bibinfo{author}{\bibfnamefont{N.}~\bibnamefont{Delone}}, \bibnamefont{and}
  \bibinfo{author}{\bibfnamefont{V.}~\bibnamefont{Krainov}},
  \bibinfo{journal}{Zh. Eksp. Teor. Fiz.} \textbf{\bibinfo{volume}{91}},
  \bibinfo{pages}{2008} (\bibinfo{year}{1986}).

\bibitem[{\citenamefont{Tong et~al.}(2002)\citenamefont{Tong, Zhao, and
  Lin}}]{Tong02}
\bibinfo{author}{\bibfnamefont{X.~M.} \bibnamefont{Tong}},
  \bibinfo{author}{\bibfnamefont{Z.~X.} \bibnamefont{Zhao}}, \bibnamefont{and}
  \bibinfo{author}{\bibfnamefont{C.~D.} \bibnamefont{Lin}},
  \bibinfo{journal}{Phys. Rev. A} \textbf{\bibinfo{volume}{66}},
  \bibinfo{pages}{033402} (\bibinfo{year}{2002}).

\bibitem[{\citenamefont{Keldysh}(1965)}]{Keldysh1965}
\bibinfo{author}{\bibfnamefont{L.}~\bibnamefont{Keldysh}},
  \bibinfo{journal}{JETP} \textbf{\bibinfo{volume}{20}}, \bibinfo{pages}{1307}
  (\bibinfo{year}{1965}).

\bibitem[{\citenamefont{Pfeiffer
  et~al.}(2012{\natexlab{b}})\citenamefont{Pfeiffer, Cirelli, Landsman,
  Smolarski, Dimitrovski, Madsen, and Keller}}]{Keller12}
\bibinfo{author}{\bibfnamefont{A.~N.} \bibnamefont{Pfeiffer}},
  \bibinfo{author}{\bibfnamefont{C.}~\bibnamefont{Cirelli}},
  \bibinfo{author}{\bibfnamefont{A.~S.} \bibnamefont{Landsman}},
  \bibinfo{author}{\bibfnamefont{M.}~\bibnamefont{Smolarski}},
  \bibinfo{author}{\bibfnamefont{D.}~\bibnamefont{Dimitrovski}},
  \bibinfo{author}{\bibfnamefont{L.~B.} \bibnamefont{Madsen}},
  \bibnamefont{and} \bibinfo{author}{\bibfnamefont{U.}~\bibnamefont{Keller}},
  \bibinfo{journal}{Phys. Rev. Lett.} \textbf{\bibinfo{volume}{109}},
  \bibinfo{pages}{083002} (\bibinfo{year}{2012}{\natexlab{b}}).

\bibitem[{\citenamefont{Budil et~al.}(1993)\citenamefont{Budil, Sali\`eres,
  L'Huillier, Ditmire, and Perry}}]{PhysRevA.48.R3437}
\bibinfo{author}{\bibfnamefont{K.~S.} \bibnamefont{Budil}},
  \bibinfo{author}{\bibfnamefont{P.}~\bibnamefont{Sali\`eres}},
  \bibinfo{author}{\bibfnamefont{A.}~\bibnamefont{L'Huillier}},
  \bibinfo{author}{\bibfnamefont{T.}~\bibnamefont{Ditmire}}, \bibnamefont{and}
  \bibinfo{author}{\bibfnamefont{M.~D.} \bibnamefont{Perry}},
  \bibinfo{journal}{Phys. Rev. A} \textbf{\bibinfo{volume}{48}},
  \bibinfo{pages}{R3437} (\bibinfo{year}{1993}).

\bibitem[{\citenamefont{Landsman et~al.}(2013)\citenamefont{Landsman, Pfeiffer,
  Hofmann, Smolarski, Cirelli, and Keller}}]{Landsman_2013}
\bibinfo{author}{\bibfnamefont{A.~S.} \bibnamefont{Landsman}},
  \bibinfo{author}{\bibfnamefont{A.~N.} \bibnamefont{Pfeiffer}},
  \bibinfo{author}{\bibfnamefont{C.}~\bibnamefont{Hofmann}},
  \bibinfo{author}{\bibfnamefont{M.}~\bibnamefont{Smolarski}},
  \bibinfo{author}{\bibfnamefont{C.}~\bibnamefont{Cirelli}}, \bibnamefont{and}
  \bibinfo{author}{\bibfnamefont{U.}~\bibnamefont{Keller}},
  \bibinfo{journal}{New J.~Phys.} \textbf{\bibinfo{volume}{15}},
  \bibinfo{pages}{013001} (\bibinfo{year}{2013}).

\end{thebibliography}
\end{document}